\documentclass[reprint,preprintnumbers,superscriptaddress,%
               amsmath,amssymb,aps,prd,longbibliography]{revtex4-2}

\usepackage{graphicx}% Include figure files
\usepackage{dcolumn}% Align table columns on decimal point
\usepackage{bm}% bold math
\usepackage[colorlinks = true,
            linkcolor = red,
            urlcolor  = blue,
            citecolor = red,
            anchorcolor = blue]{hyperref}% add hypertext capabilities
\usepackage[capitalise]{cleveref}
\usepackage[caption=false]{subfig}
\usepackage{multirow}
\usepackage{siunitx}
\usepackage{subfig} 
\usepackage[mathlines]{lineno}% Enable numbering of text and display math
\begin{document}

\preprint{FERMILAB-PUB-26-0632-PPD}
\title{Search for Neutrino-Produced Long-Lived Particles Using the SBND Cosmic Ray Tagger Beam Telescope}% Force line breaks with \\
%\title{SBND Interim Design Report (IDR) Author List - alphabetical - August 2026}

%----------------------------------------------------
% List of institutions, need not be alphabetical
% Need to also add institution as \affiliation below this to get alphabetical order right
\newcommand{\ABC}{Universidade Federal do ABC, Santo Andre, SP 09210-580, Brazil}
\newcommand{\ANL}{Argonne National Laboratory, Lemont, IL 60439, USA}
\newcommand{\Alfenas}{Universidade Federal de Alfenas, Pocos de Caldas, MG 37715-400, Brazil}
\newcommand{\BNL}{Brookhaven National Laboratory, Upton, NY 11973, USA}
\newcommand{\Bern}{Universit{\"a}t Bern, Bern CH-3012, Switzerland}
\newcommand{\CIEMAT}{Centro de Investigaciones Energ{\'e}ticas, Medioambientales y Tecnol{\'o}gicas, Madrid E-28040, Spain}
\newcommand{\CSU}{Colorado State University, Fort Collins, CO 80523, USA}
\newcommand{\CTI}{Centro de Tecnologia da Informa{\c{c}}{\~a}o Renato Archer, Campinas, SP 13069-901, Brazil}
\newcommand{\Campinas}{Universidade Estadual de Campinas, Campinas, SP 13083-970, Brazil}
\newcommand{\Chicago}{University of Chicago, Chicago, IL 60637, USA}
\newcommand{\Columbia}{Columbia University, New York, NY 10027, USA}
\newcommand{\Edinburgh}{University of Edinburgh, Edinburgh EH9 3FD, United Kingdom}
\newcommand{\FNAL}{Fermi National Accelerator Laboratory, Batavia, IL 60510, USA}
\newcommand{\Florida}{University of Florida, Gainesville, FL 32611, USA}
\newcommand{\GeorgiaTech}{Georgia Institute of Technology, Atlanta, GA 30332, USA}
\newcommand{\Granada}{Universidad de Granada, Granada E-18071, Spain}
\newcommand{\Hanyang}{Hanyang University, 222, Wangsimni-ro, Seongdong-gu, Seoul, 04763, Korea}
\newcommand{\Holyoke}{Mount Holyoke College, South Hadley, MA 01075, USA}
\newcommand{\IIS}{Indian Institute of Science, Centre for High Energy Physics, Bangalore 560012, India}
\newcommand{\IIT}{Illinois Institute of Technology, Chicago, IL 60616, USA}
\newcommand{\Imperial}{Imperial College London, London SW7 2AZ, United Kingdom}
\newcommand{\Kansas}{University of Kansas, Lawrence, KS 66045, USA}
\newcommand{\LANL}{Los Alamos National Laboratory, Los Alamos, NM 87545, USA}
\newcommand{\LSU}{Louisiana State University, Baton Rouge, LA 70803, USA}
\newcommand{\Lancaster}{Lancaster University, Lancaster LA1 4YW, United Kingdom}
\newcommand{\Liverpool}{University of Liverpool, Liverpool L69 7ZE, United Kingdom}
\newcommand{\Manchester}{University of Manchester, Manchester M13 9PL, United Kingdom}
\newcommand{\Michigan}{University of Michigan, Ann Arbor, MI 48109, USA}
\newcommand{\Minnesota}{University of Minnesota, Minneapolis, MN 55455, USA}
\newcommand{\NotreDame}{University of Notre Dame, Notre Dame, IN 46556, USA}
\newcommand{\Oxford}{University of Oxford, Oxford OX1 3RH, United Kingdom}
\newcommand{\Palermo}{Universit{\`a} degli Studi di Palermo, Palermo I-90123, Italy}
\newcommand{\Penn}{University of Pennsylvania, Philadelphia, PA 19104, USA}
\newcommand{\QueenMary}{Queen Mary University of London, London E1 4NS, United Kingdom}
\newcommand{\Rutgers}{Rutgers University, Piscataway, NJ 08854, USA}
\newcommand{\SaoJose}{Instituto Tecnol{\'o}gico de Aeron{\'a}utica S{\~a}o Jos{\'e} dos Campos, SP 12228-900, Brazil}
\newcommand{\Sheffield}{University of Sheffield, School of Mathematical and Physical Sciences, Sheffield S3 7RH, United Kingdom}
\newcommand{\Sussex}{University of Sussex, Brighton BN1 9RH, United Kingdom}
\newcommand{\Syracuse}{Syracuse University, Syracuse, NY 13244, USA}
\newcommand{\TAMU}{Texas A{\&}M University, College Station, TX 77843, USA}
\newcommand{\Tufts}{Tufts University, Medford, MA 02155, USA}
\newcommand{\UCSB}{University of California Santa Barbara, Santa Barbara, CA 93106, USA}
\newcommand{\UTA}{University of Texas at Arlington, Arlington, TX 76019, USA}
\newcommand{\VirginiaTech}{Virginia Tech, Center for Neutrino Physics, Blacksburg, VA 24060, USA}
\newcommand{\Warwick}{University of Warwick, Coventry CV4 7AL, United Kingdom}
%----------------------------------------------------
% So that institutions appear in alphabetical order:
\affiliation{\ABC}
\affiliation{\ANL}
\affiliation{\Alfenas}
\affiliation{\BNL}
\affiliation{\Bern}
\affiliation{\CIEMAT}
\affiliation{\CSU}
\affiliation{\CTI}
\affiliation{\Campinas}
\affiliation{\Chicago}
\affiliation{\Columbia}
\affiliation{\Edinburgh}
\affiliation{\FNAL}
\affiliation{\Florida}
\affiliation{\GeorgiaTech}
\affiliation{\Granada}
\affiliation{\Hanyang}
\affiliation{\Holyoke}
\affiliation{\IIS}
\affiliation{\IIT}
\affiliation{\Imperial}
\affiliation{\Kansas}
\affiliation{\LANL}
\affiliation{\LSU}
\affiliation{\Lancaster}
\affiliation{\Liverpool}
\affiliation{\Manchester}
\affiliation{\Michigan}
\affiliation{\Minnesota}
\affiliation{\NotreDame}
\affiliation{\Oxford}
\affiliation{\Palermo}
\affiliation{\Penn}
\affiliation{\QueenMary}
\affiliation{\Rutgers}
\affiliation{\SaoJose}
\affiliation{\Sheffield}
\affiliation{\Sussex}
\affiliation{\Syracuse}
\affiliation{\TAMU}
\affiliation{\Tufts}
\affiliation{\UCSB}
\affiliation{\UTA}
\affiliation{\VirginiaTech}
\affiliation{\Warwick}
%----------------------------------------------------
% Authors in alphabetical order
\author{R.~Acciarri} \affiliation{\FNAL}
\author{L.~Aliaga-Soplin} \affiliation{\UTA}
\author{D.~Andrade Aldana} \affiliation{\IIT}
\author{C.~Andreopoulos} \affiliation{\Liverpool}
\author{A.~Antonakis} \affiliation{\UCSB}
\author{S.~Balasubramanian} \affiliation{\Holyoke}
\author{A.~Barnard} \affiliation{\Oxford}
\author{V.~Basque} \affiliation{\Edinburgh}
\author{J.\,R.~Bateman} \affiliation{\Manchester}\affiliation{\Imperial}
\author{A.~Beever} \affiliation{\Sheffield}
% ...
%188 authors
%----------------------------------------------------
\collaboration{The SBND Collaboration}
\thanks{sbnd\_info@fnal.gov}%

\begin{abstract}
The Short-Baseline Near Detector (SBND) is a liquid argon time projection chamber in Fermilab’s Short-Baseline Neutrino Program, positioned 110 meters from the neutrino target along the Booster Neutrino Beam (BNB).  In addition to its neutrino physics goals, SBND enables searches for new physics, including long-lived particles (LLPs) produced via neutrino–nucleus upscattering that decay into visible final states. Although neutrino data taking with the full SBND experiment began in December 2024, a subset of the Cosmic Ray Tagger (CRT) system operated from 2017--2018 as a standalone detector, referred to as the CRT Beam Telescope, collecting data corresponding to $2.95\times 10^{20}$ protons on target from the BNB. Originally intended for commissioning and performance studies, the substantial dataset collected also enables searches for displaced LLP decays. This paper presents the first search for physics beyond the Standard Model using data from the SBND CRT Beam Telescope, targeting LLP decays into electron–positron pairs. We observe no significant excess above the expected background and set 90\% confidence-level upper limits on an effective production cross section at the level of $10^{-38}$ cm$^2$ for laboratory decay lengths of 0.3–100~m. These results complement existing searches by probing displaced decay signatures at meter-scale decay lengths.
\end{abstract}

\maketitle
\tableofcontents

%################################################
\section{Introduction}
%################################################

The Short-Baseline Near Detector (SBND)~\cite{sbnd:2026detector} is a 112 ton liquid argon time projection chamber (LArTPC) positioned 110~m downstream of the target along the Booster Neutrino Beam (BNB)~\cite{miniboone:2009flux} at Fermi National Accelerator Laboratory (Fermilab). As the near detector of the Short-Baseline Neutrino (SBN) program~\cite{MicroBooNE:2015bmn}, SBND contributes to a definitive multi-detector search for sterile neutrino oscillations at the eV mass scale and other phenomena beyond the Standard Model (BSM). SBND in particular offers a rich BSM physics program, enabled by the intense neutrino flux of the BNB and the detector's proximity to the neutrino source~\cite{Machado:2019oxb}.

In BSM theories, it is common for one or more newly introduced particles to have sufficiently long lifetimes to travel macroscopic distances before decaying. Searches for such particles, and their subsequent decays into observable Standard Model (SM) final states, fall under the moniker of long-lived-particle (LLP) searches. Conventionally, these searches at accelerator neutrino experiments focus on LLPs produced through meson decays~\cite{MicroBooNE:2023eef,MicroBooNE:2022ctm, MicroBooNE:2023gmv,T2K:2019jwa,ICARUS:2024oqb}. However, intense neutrino beams, such as the BNB, offer a complementary production environment: LLPs can be generated via neutrino–nucleus scattering in the surrounding material and subsequently decay into visible final states downstream. The analysis presented here therefore targets LLPs produced through neutrino interactions, rather than the conventional meson-decay channel.

This paper presents a search for LLPs using data from the SBND experiment. The analysis is performed using a standalone configuration of the SBND Cosmic Ray Tagger (CRT) system, referred to as the CRT Beam Telescope. Prior to the installation of the SBND cryostat, which houses the LArTPC, a subset of CRT modules was deployed in the detector hall and operated between 2017 and 2018, as shown in~\cref{fig:CRTBT}. The detector collected data corresponding to $2.95\times 10^{20}$ protons on target (POT) from the BNB. The CRT Beam Telescope was originally intended for commissioning and performance studies, and not as a dedicated BSM detector. Despite this, its geometry, spatial/timing resolution, and exposure to the neutrino beam have been shown to provide effective background rejection enabling a search for displaced decays at macroscopic distances from a LLP’s production point.

The LLPs considered in this analysis are produced via neutrino upscattering in the material, approximated as silicon, between the BNB target and the CRT Beam Telescope, outside the detector volume. They are modeled to propagate macroscopic distances before decaying into electron–positron pairs. As a result, the signal originates from displaced decays occurring within the volume between the two CRT panels, rather than from interactions within instrumented material. This distinguishes the present search from conventional neutrino interaction measurements, which target prompt, in-detector vertices. This search sets limits on the product of the LLP production cross section via neutrino upscattering and the LLP decay branching ratio, as a function of LLP lifetime. This parameter space has not previously been explored by other experiments.

To guide the signal modeling, this search is inspired by dark neutrino scenarios~\cite{Bertuzzo:2018itn,Ballett:2019pyw}, in which a SM neutrino upscatters into a heavier neutral fermion that subsequently decays into visible final states. Rather than focusing on one particular model realization, the results are presented in a framework that parametrizes the LLP production rate and decay lengths, allowing reinterpretation for models with similar phenomenology, for example when additional final states or dark neutrinos are added to the model that can modify the LLP lifetime substantially.

This paper is organized as follows. \cref{sec:detector} describes the CRT Beam Telescope detector configuration and the beamline. \cref{sec:llp} introduces the LLP model assumptions and the signal simulation. The background estimation is presented in \cref{sec:Background}. \cref{sec:Reconstruction} describes the reconstruction framework and its performance. The analysis strategy, event selection, and signal–background discrimination are presented in \cref{sec:selection}, and systematic uncertainties are discussed in \cref{sec:systematics}. The results and their interpretation are given in \cref{sec:results}, and the conclusions are summarized in \cref{sec:conclusions}.

%################################################
\section{The CRT Beam Telescope and the BNB}
\label{sec:detector}
%################################################

Operating at the surface, SBND is instrumented with plastic scintillator panels forming the CRT~\cite{Auger:2016tjc} system, which tags and rejects cosmic backgrounds. A total of twenty-four CRT modules were installed in the SBND detector hall in 2017, prior to the installation of the cryostat.

The basic unit of the CRT is a polystyrene scintillator strip, in which charged particles produce scintillation photons. The light is collected by two wavelength-shifting fibers glued into grooves along the long edges of the strip and guided to its readout end, where each fiber is coupled to a Hamamatsu S12825-050P silicon photomultiplier (SiPM)~\cite{Hamamatsu}. The relative signal amplitudes of the SiPM pairs provide position information within the module. Each module is read out by a dedicated front-end board that biases, amplifies, and digitizes the SiPM signals and provides event timestamps with $\sim$1~ns precision; further details of the design are given in~\cite{Auger:2016tjc,Auger:2016vpo}.

\begin{figure}[!htbp]
    \centering
    \includegraphics[width=\linewidth]{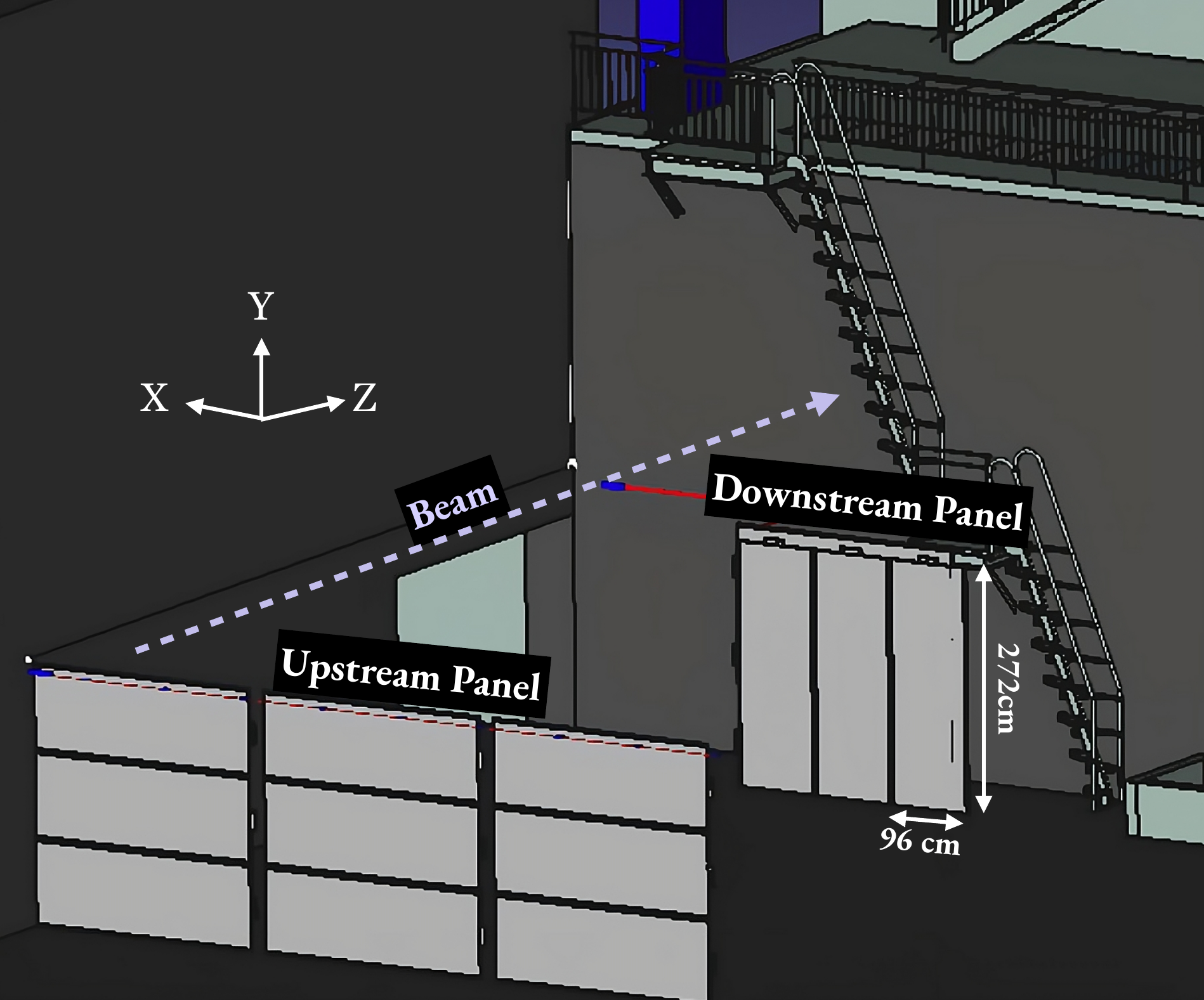}
    \caption{\label{fig:CRTBT} Illustration of the CRT Beam Telescope detector setup, showing the size of a CRT module, the upstream and downstream panels separated by 9.26~m along the beam direction, and the coordinate system. Modules within each panel are arranged in two layers with horizontal and vertical strip direction; coincident hits in perpendicular modules provide two-dimensional position reconstruction.}
\end{figure}

Each CRT module consists of sixteen strips and measures approximately 272~cm in length, 96~cm in width, and 2~cm in thickness. The modules are grouped into two panels oriented perpendicular to the beam direction and separated by 9.26~m, forming the upstream and downstream panels of the SBND CRT Beam Telescope (\cref{fig:CRTBT}). Each panel contains both horizontally and vertically oriented modules; a coincidence is formed when overlapping horizontal and vertical modules register signals within a narrow time window ($\sim$20~ns), providing a two-dimensional hit position. With a timing resolution of $\sim$1.5~ns, the CRT enables precise temporal coincidence measurements between the upstream and downstream panels.

The upstream panel (approximately 23.5~m$^2$) consists of eighteen modules, while the downstream panel (approximately 7.8~m$^2$) consists of six modules. Both panels are centered with respect to each other along the beam direction. The CRT Beam Telescope uses a right-handed coordinate system where the $x$-axis is perpendicular to the neutrino beam direction, the $y$-axis is vertical, and the $z$-axis follows the beam direction.

The hypothetical LLPs and other particles detectable by the CRT Beam Telescope are produced using the BNB. The BNB at Fermilab produces neutrinos by colliding protons with a kinetic energy of 8 GeV into a beryllium target~\cite{miniboone:2009flux}. Secondary mesons produced in the collisions, primarily pions and kaons, are focused (or defocused) by the electromagnetic horn depending on their charge polarity. When positive mesons are focused with the horn, before entering and decaying in a 45~m long pipe, this gives rise to a neutrino-enhanced beam, referred to as neutrino mode. A 4~m thick steel absorber is placed at the end of the decay region to stop most remaining particles, with the exception of neutrinos. In neutrino mode, the beam composition is approximately $\sim$93.5\% $\nu_\mu$, $\sim$6\% $\bar{\nu}_\mu$, $\sim$0.45\% $\nu_e$, and $\sim$0.05\% $\bar{\nu}_e$~\cite{SBND:2025plf}, with neutrino energies peaking at approximately 800~MeV. The BNB runs 7~m below ground and parallel to the Earth’s surface. After the absorber, neutrinos produced in the beamline must travel an additional 60~m through the ground to reach the SBND detector building. 

The beam is delivered in pulses of 1.6~$\mu$s duration with a rate of $\sim$5 Hz, and a typical intensity of $\sim$5$\times$10$^{12}$ protons per pulse. The data used in this analysis were collected in neutrino mode.

%################################################
\section{LLP Signal Modeling and Simulation}
\label{sec:llp}
%################################################

\begin{figure}[!htbp]
    \centering
    \includegraphics[width=.6\linewidth]{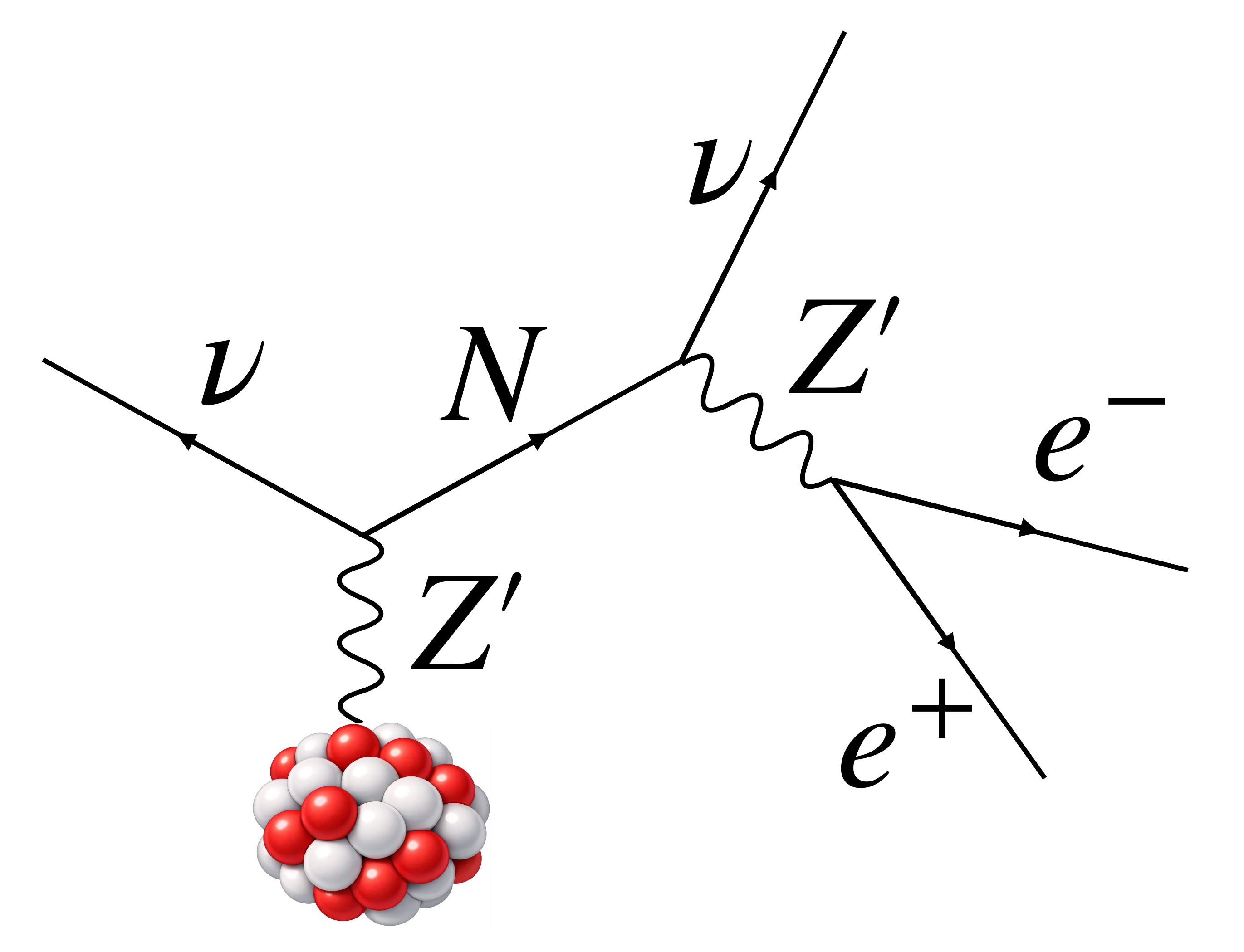} 
    \caption{\label{fig:SignalIllustartion} Illustration of the benchmark LLP production via neutrino upscattering and subsequent decay. A SM neutrino $\nu$ from the BNB upscatters off a nuclear target, producing an LLP, $N$, which propagates a macroscopic distance before decaying into $e^+e^-\nu$. }
\end{figure}

As described in the previous section, the roughly ten-meter separation between the detector panels defines a decay volume. As such, the CRT Beam Telescope is naturally suited to search for LLPs that are produced upstream and decay in flight between the panels. We focus on a specific subclass of LLP searches: scenarios in which an SM neutrino upscatters in the material upstream of the detector into a BSM particle state. This exotic particle will then subsequently decay with a macroscopic decay length. This phenomenology is predicted in ``dark neutrino'' models and many associated scenarios~\cite{Bertuzzo:2018itn,Ballett:2018ynz,Ballett:2019pyw,Abdullahi:2020nyr,Dutta:2025npn}. A diagram demonstrating this two-step process is provided in \cref{fig:SignalIllustartion}.

As a benchmark, we consider an LLP signal with the following parameters:
\begin{align}
    m_{N} = 400\,\mathrm{MeV},\quad m_{Z^\prime} &= 1.25\,\mathrm{GeV}, \nonumber \\
    \left\lvert U_{\mu N}\right\rvert^2 = 4 \times 10^{-6},\quad \varepsilon &= 2 \times 10^{-2}.
\end{align}
The parameter $U_{\mu N}$ denotes the coupling strength between the SM muon neutrino and the LLP, and $\varepsilon$ parametrizes the kinetic mixing of the dark mediator $Z^\prime$. These parameter values define the signal kinematics and detector acceptance for this benchmark scenario. For this set of parameters, the inclusive cross section for an SM neutrino to upscatter into the LLP $N$ in the surrounding material, including both coherent scattering off nuclei and incoherent elastic scattering off nucleons, is approximately $2 \times 10^{-38}~\mathrm{cm}^2$, provided $E_\nu \gtrsim 1~\mathrm{GeV}$ — the energy range that dominates LLP production which is boosted enough to be detected in our detector. With these coupling strengths, the LLP decay length $c\tau$ is calculated to be 3.1~m, after which the LLP decays off-shell into the $e^+e^-\nu$ final state.

In this work we focus on fixed values of $m_N$ and $m_{Z^\prime}$, while allowing the inclusive upscattering cross section $\sigma(\nu A \to NX)$, where $A$ and $X$ are the target nucleus before and after scattering, and the LLP decay length to vary as free parameters. Because we focus on decays of the LLP into electron-positron pairs and our detector is insensitive to neutrinos, we are only sensitive to the product of the production cross section and the branching ratio for $N \to e^+e^- + E_{\mathrm{miss}}$, $\sigma(\nu A \to NX) \times \mathrm{Br}(N \to e^+e^- + E_{\mathrm{miss}})$.

This approach allows reinterpretation for BSM scenarios sharing similar production and decay phenomenology. 
This includes scenarios with multiple dark neutrinos, such as those in which a heavier state $N_5$ decays as $N_5 \to N_4 e^+e^-$, provided that $N_4$ is sufficiently light that the visible decay kinematics and detector acceptance remain similar to those of the benchmark considered here. Dedicated \texttt{DarkNews} studies indicate that, for $m_N \lesssim 600~\text{MeV}$, the neutrino-energy dependence of the upscattering cross section is only weakly affected by the choice of $m_N$ once above the production threshold, supporting the applicability of these limits over this mass range. The final results are presented as exclusion limits in the plane of LLP decay length versus effective production rate.

    \subsection{Simulation Details}
    The LLP production is performed using the \texttt{DarkNews} package~\cite{Abdullahi:2022cdw}, which focuses on specific model-based scenarios in the BSM parameter space. In what follows, to ensure a more generic analysis, we allow the LLP decay length and its production cross section to vary independently. We find that as long as the LLP mass is small relative to the BNB neutrino energy ($\mathcal{O}(100)$~MeV) and the BSM mediator $Z^\prime$ is relatively heavy ($\gtrsim$~GeV), this generic approach applies broadly.

    \texttt{DarkNews} generates events using a set of neutrino fluxes by flavor provided as a one-dimensional distribution in neutrino energy. Owing to the large transverse extent of the CRT Beam Telescope and the collimated nature of the beam, the neutrino flux varies significantly across the detector surface. This well-known off-axis effect~\cite{SBND:2025plf} is accounted for by generating multiple off-axis neutrino flux distributions from a full flux simulation~\cite{miniboone:2009flux}, adapted to the detector location along the beamline and provided as inputs to \texttt{DarkNews} for event generation.
    
    Signal samples with different LLP decay lengths are generated by applying event-by-event weights to the benchmark model with an unboosted decay length ${c\tau = 3.1~\mathrm{m}}$. For each decay length considered, the scaling factor used to reweight each simulated event is calculated as:
    
    \begin{equation}
        r = \frac{l^{-1}\exp(-l'/\gamma\beta\, l) }{l_{\text{benchmark}}^{-1}\exp(-l'/\gamma\beta\, l_{\text{benchmark}})}, 
        \label{eq:signalweight}
    \end{equation}
    where $\gamma = E_N/m_N$, $\beta = \sqrt{1 - 1/\gamma^2}$, $l$ is the tested decay length in the rest frame, $l^{\prime}$ is the tested decay length in the lab frame, and $l_{\text{benchmark}}$ is the decay length in the rest frame in the benchmark model. The decay lengths are chosen to span the region of parameter space in which a decay can occur within the geometric acceptance of the CRT Beam Telescope. Six signal samples are generated, with decay lengths of 0.31, 1, 3.1, 10, 50, and 100~m.

%################################################  
\section{Background Characterization}
\label{sec:Background}
%################################################

The dominant backgrounds for this analysis come from neutrino interactions in the soil and building material surrounding the detector, and from cosmogenic activity. Neutrinos interacting around the CRT panels can produce secondary particles that then interact with the CRT, mimicking an LLP signature. We simulate neutrino interactions on surrounding materials of the SBND detector hall using the \texttt{GENIE} neutrino event generator (\texttt{v3.00.06p G1810a0211a})~\cite{Andreopoulos:2009rq, PhysRevD.105.072001}. A total of approximately $2.78 \times 10^{21}$ POT equivalent neutrino interactions, corresponding to 9 times the total beam telescope dataset statistics, were simulated to model the expected background. These simulated interactions are used to estimate neutrino-induced backgrounds relevant for this analysis.

The propagation of simulated particles, as well as the energy deposition in the scintillators, are simulated using the \texttt{GEANT4} particle propagation toolkit (version \texttt{v4.10.6.p01d})~\cite{GEANT4:2002zbu} with the \texttt{QGSP\_BERT\_HP} physics list~\cite{GEANT4:PhysicsManual}. The deposited energy simulated by \texttt{GEANT4} is then passed to a custom CRT Beam Telescope detector simulation within \texttt{sbndcode v09\_59\_00}~\cite{sbndcode} to model the charge response and timing. Once the charge signal exceeds a predefined threshold and satisfies the coincidence requirement between vertical and horizontal modules (as described in~\cref{sec:detector}), a trigger is simulated and events with the corresponding waveform amplitude and timing information are recorded, thus mimicking real data acquisition. This particle transport and detector simulation is also applied identically to our simulated signal events.

The second dominant background arises from cosmogenic activity. Rather than relying on simulation, this background is estimated directly from data using an independent sample. The CRT data acquisition window is 1~s in duration, which is much larger than the 1.6~$\mu$s BNB spill window. Since LLP decays are expected to occur either approximately in temporal coincidence, or after the BNB spill, a 1.6~$\mu$s time window immediately preceding the neutrino beam spill is selected as a background-only region, as it is free from both neutrino-induced interactions and potential LLP signals. CRT activity measured in this pre-spill window is used to estimate both the cosmogenic background and the accidental cosmogenic veto rate for signal events, the latter of which is found to be negligible and has a minimal impact on the signal efficiency. The cosmogenic background is statistically subtracted from all distributions presented in this paper. This approach provides a fully data-driven determination of the cosmic-ray background, avoiding any dependence on external simulations.

%################################################    
\section{Event Reconstruction}
\label{sec:Reconstruction}
%################################################

This section describes how data and simulated events are reconstructed for use in the downstream analysis.

A \textit{CRT hit} is formed by combining signals from two strips of different orientation (vertical and horizontal), and is characterized by its time of arrival and signal amplitude, from which timing, position, and deposited energy are subsequently derived. The position within each strip is determined by weighting according to the charge ratio measured by the two SiPMs. The overlapping geometry of the vertical and horizontal modules can produce combinatorial ambiguities when forming hit candidates from multiple coincident strip signals. Several horizontal–vertical module combinations may therefore be reconstructed, where only one corresponds to the true particle crossing. This effect is intrinsic to the detector geometry and increases with strip multiplicity. Identifying the correct combination is challenging, and an incorrect pair may be selected; such misreconstructed candidates are referred to as ``fake hits.'' The selection of a fake hit in place of the true one, however, has little impact on the downstream analysis, which is largely insensitive to the specific pairing chosen. 

\begin{figure}[!htbp]
    \centering    
    \includegraphics[width=\linewidth]{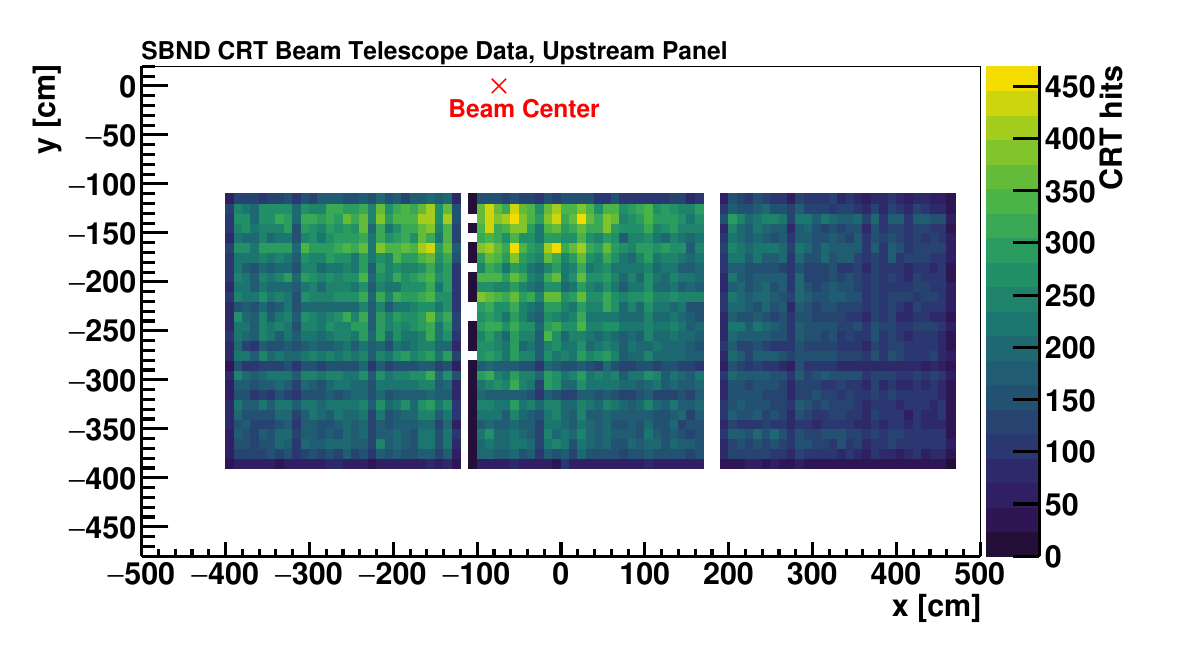}   
    \label{fig:BeamCenter} 
    \caption{CRT hit positions in the $x$–$y$ plane for the CRT Beam Telescope upstream panel during beam operation ($2.95\times10^{20}$ POT). The beam profile is clearly visible, with the beam center located at (-74, 0) cm [red cross], approximately 100 cm above the detector. Hits result from charged particles produced in neutrino interactions with surrounding material traversing the CRT modules.} 
    \label{fig:CRTdata}
\end{figure}

After the position is reconstructed, corrections for both charge and timing are applied to account for detector effects. The reconstructed CRT hit positions in the $x$--$y$ plane for the upstream panel are shown in \cref{fig:CRTdata}, where a clear beam profile is visible. We note that the distribution of hits in the downstream panel is more spatially uniform due to the angular spread of charged particles produced by neutrinos interacting with the surrounding soil and concrete, and the fact that the downstream panel detects particles from interactions that are on average farther away from the panel than those interacting with the upstream panel. Moreover, as illustrated in \cref{fig:CRTdata}, the center of the beamline is located approximately 100~cm above the top of the CRT Beam Telescope detector setup. This results in an off-axis configuration where the detector observes neutrinos from approximately $0.5^\circ$ to $2.5^\circ$ relative to the beam direction as measured from the target. Such detector position relative to the neutrino beam determines the geometric acceptance for LLP searches.

\begin{figure}[!htbp]
  \centering
  \includegraphics[width=0.9\linewidth]{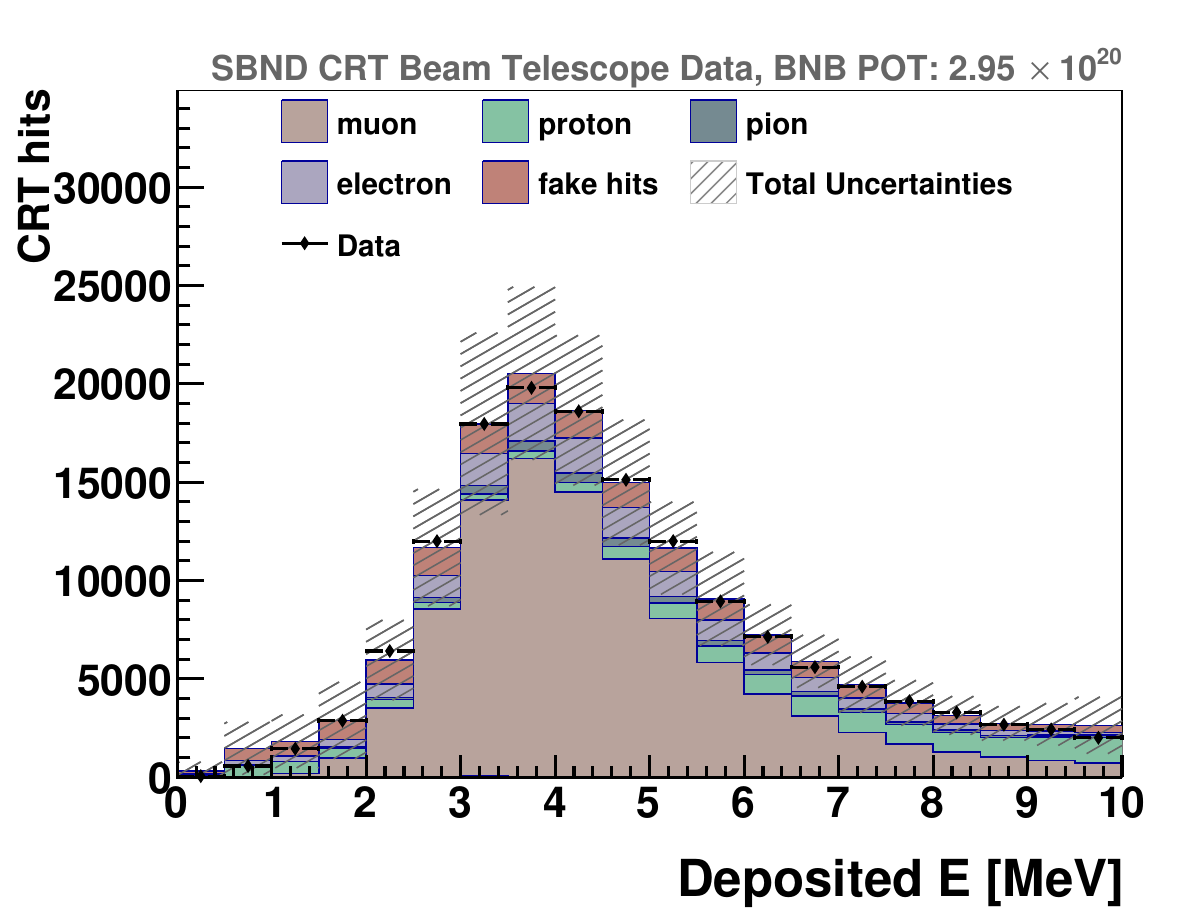}
  \caption{Distribution of deposited energy in the downstream CRT panel, comparing data ($2.95\times10^{20}$ POT) with Monte Carlo predictions broken down by particle type: muon, proton, pion, and electron. Muons dominate the spectrum and exhibit minimum-ionizing behavior, while protons deposit more energy. ``Fake hits" arise from combinatorial ambiguities in the reconstruction of multiple coincident strips.}
 \label{fig:depositedE_newPhysicsList}
\end{figure}

The deposited energy is then evaluated from the calibrated hit amplitudes, with calibration applied to account for detector effects and ensure consistency between data and simulation. Calibration uses crossing tracks, corrected for incident angle, to recalibrate the energy loss per unit length from simulation to data. The resulting distribution for the downstream panel is shown in \cref{fig:depositedE_newPhysicsList}, comparing data to the simulated neutrino background broken down by particle type. Beam-induced muons from neutrino interactions in the surrounding soil are the dominant background and exhibit the characteristic minimum-ionizing behavior, while protons deposit considerably more energy. The CRT provides sufficient separation between minimum-ionizing particles and protons for the purposes of this analysis.

Reconstruction performance has been validated using a sample of cosmogenic muons, which provide well-understood minimum-ionizing tracks traversing both CRT panels, and are used to verify the timing and position reconstruction.

%################################################
\section{Event Selection and Background Discrimination}
\label{sec:selection}
%################################################

\begin{figure}[!htbp]
    \centering
    \includegraphics[width=\linewidth]{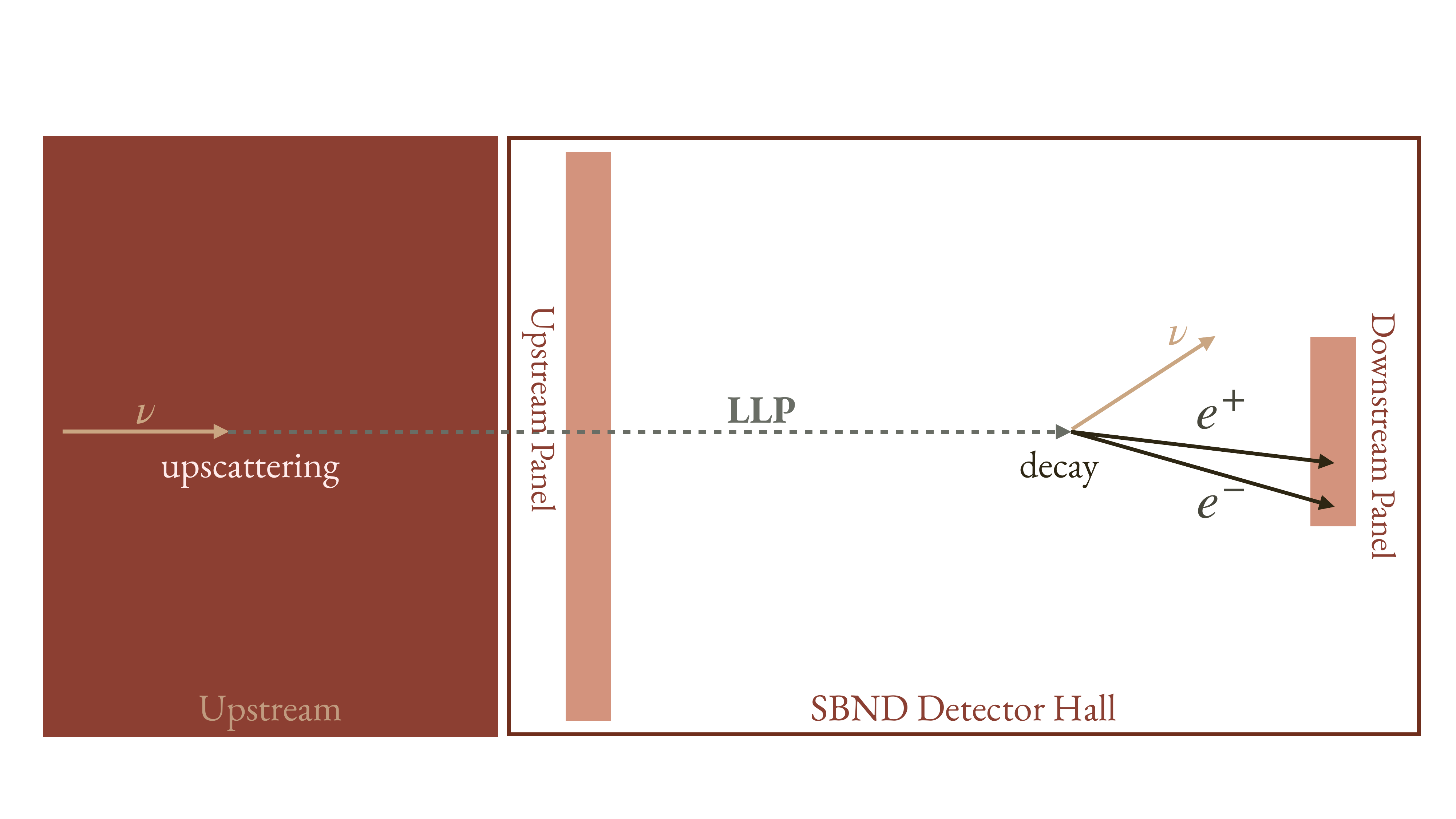}
    \caption{Schematic side view of the expected LLP signal topology in the CRT Beam Telescope geometry. A neutrino ($\nu$) from the BNB upscatters in the material upstream of the detector hall, producing an LLP that travels a macroscopic distance before decaying into $\nu, e^+$, and $e^-$. The $e^+$ and $e^-$ are detected in the downstream CRT panel (right). The upstream panel (left) registers no activity, providing background rejection. }
    \label{fig:SignalInCRTBT} 
\end{figure}

The signal, illustrated in \cref{fig:SignalInCRTBT}, consists of an LLP produced via neutrino upscattering in material upstream of the detector, which subsequently decays into an $e^+e^-$ pair. We require no activity in the upstream panel, which serves as a veto against neutrino-induced background. This requirement has minimal impact on signal efficiency — signal LLPs are defined as those decaying between the two panels, whose decay products do not traverse the upstream panel — while significantly reducing neutrino-induced background. We refer to this requirement as the \emph{Upstream Veto} cut.

Monte Carlo simulation of our benchmark signal shows that 89\% of events consist of a single hit in the downstream panel, 8\% of two hits, and 3\% of more than two hits. This indicates that in most cases only one of the outgoing leptons intersects the downstream CRT, reflecting the interplay between the decay kinematics and the detector geometry. For a 400~MeV LLP, the distribution of the $e^{+}e^{-}$ opening angle peaks at $33^\circ$. This relatively large opening angle, combined with the $\sim$100~cm vertical offset between the beam axis and the center of the CRT panels (\cref{fig:CRTdata}), results in most events having only one decay product intercepting the downstream panel. While single-hit events comprise the majority of the signal events, they are largely overwhelmed by the beam-induced muon background, which, as can be seen in~\cref{fig:depositedE_newPhysicsList}, dominates all energy depositions when no constraints are applied. Requiring two hits substantially suppresses this background, and as such defines the second cut used in this analysis (\emph{Two-Hit Requirement}). The remaining 3.0\% of events arise from additional $e^{\pm}$ interactions in the CRT that produce extra hits, and are excluded from this analysis.

Finally, motivated by comparing the hit separation distance distributions for simulated signal and background, which reveals a large excess below 22~cm arising from neutrino interactions occurring inside the CRT itself and producing activity in adjacent CRT strips, we apply a \emph{Minimum Separation} cut, requiring a distance greater than 22~cm between the two selected hits. 

These three cuts --- \emph{Upstream Veto}, \emph{Two-Hits requirement}, and \emph{Minimum Separation} --- form the minimal selection criteria to remove the most obvious neutrino-induced backgrounds. Since the cosmogenic background is measured directly from data and is statistically subtracted, the remaining dominant background is neutrino interaction events, which the analysis focuses on rejecting. According to our benchmark model simulation, the minimal selection cuts reject approximately 99.86\% of the simulated background sample, leaving 767 background events and 16 signal events for $2.95\times10^{20}$ POT. While the background rejection is high, the purity of our sample is still only 2\%, necessitating further background discrimination. This is achieved through a multivariate discrimination method using a Boosted Decision Tree (BDT).

The BDT is trained on position, timing, and deposited energy information from the CRT, using the Toolkit for Multivariate Data Analysis (TMVA) package~\cite{hoecker2009tmvatoolkitmultivariate}. A separate BDT is trained for each LLP decay length considered in the analysis, using 50\% of the available signal and background statistics, with the remaining 50\% reserved for statistical inference and final result extraction. The following variables are used as input to the BDT:
\begin{itemize}
    \item The distance between the two hits.
    \item The horizontal distance between the two hits.
    \item The timing difference between the two hits.
    \item The mean deposited energy of the two hits.
\end{itemize}

\begin{figure*}[!htbp]
    \centering
    \subfloat[]{
        \includegraphics[width=0.34\linewidth]{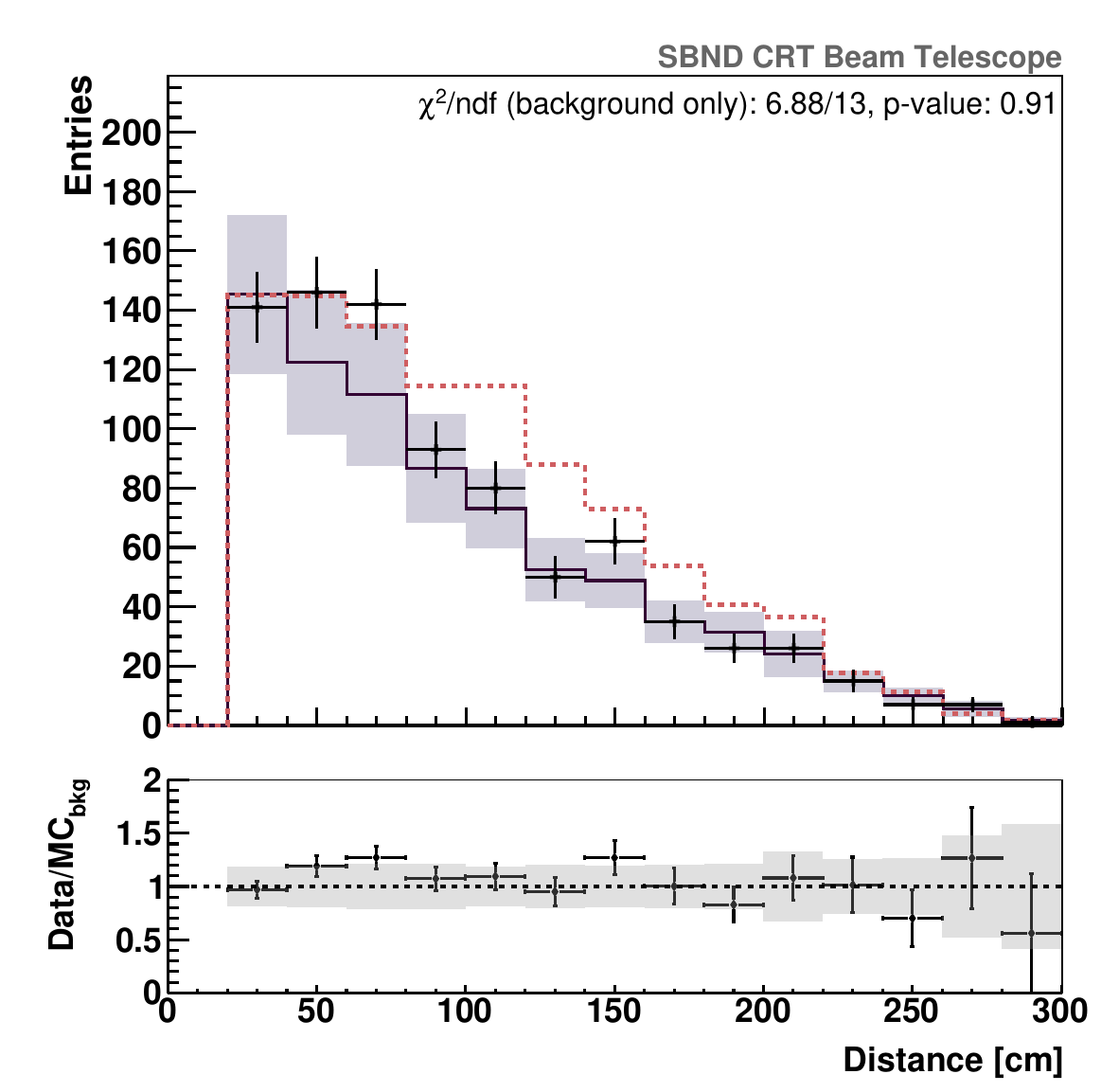}
         \label{fig:BDTVars_a}}
    \subfloat[]{
        \includegraphics[width=0.34\linewidth]{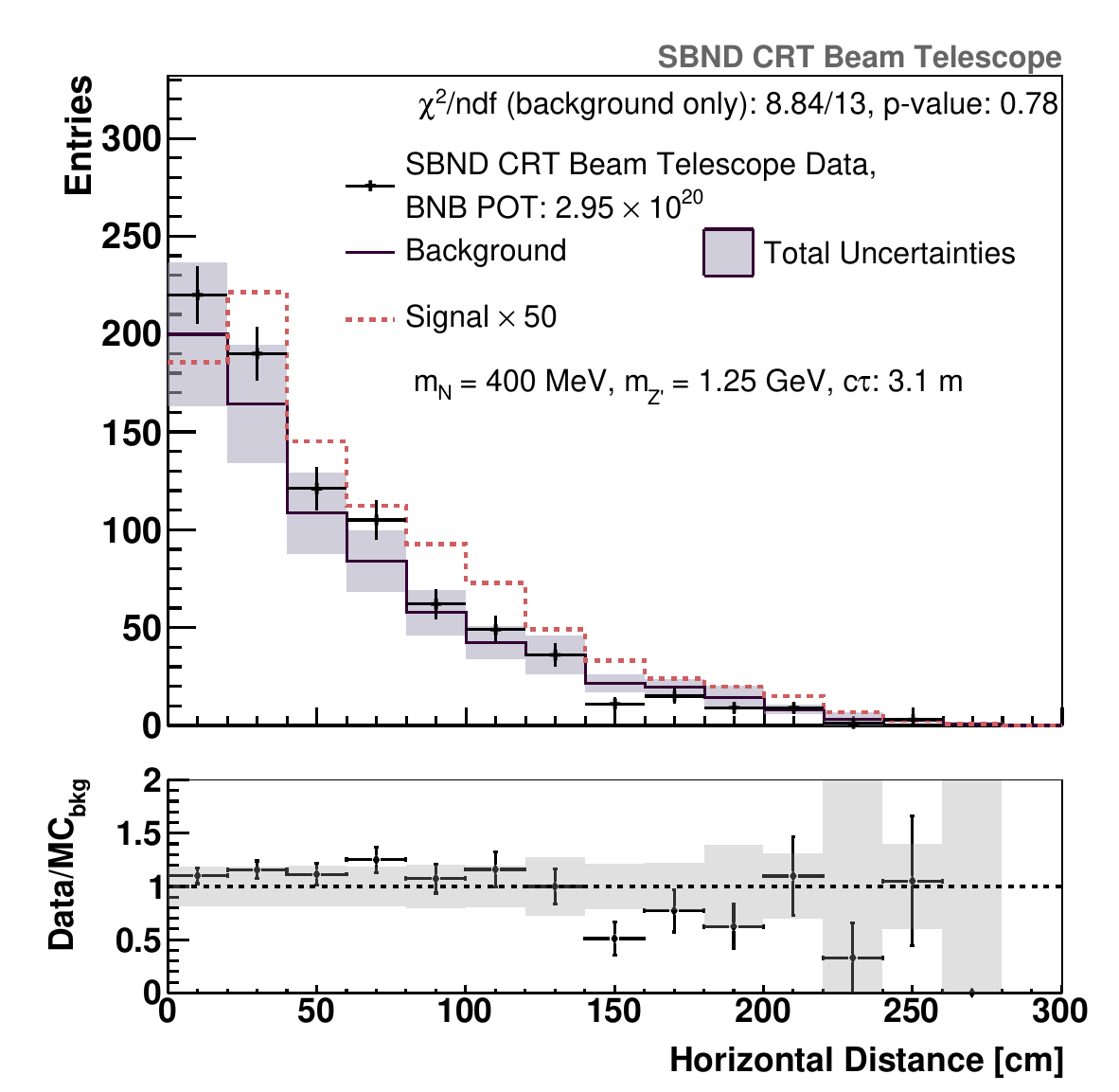}
        \label{fig:BDTVars_b}}
    \hfill
    \subfloat[]{
        \includegraphics[width=0.34\linewidth]{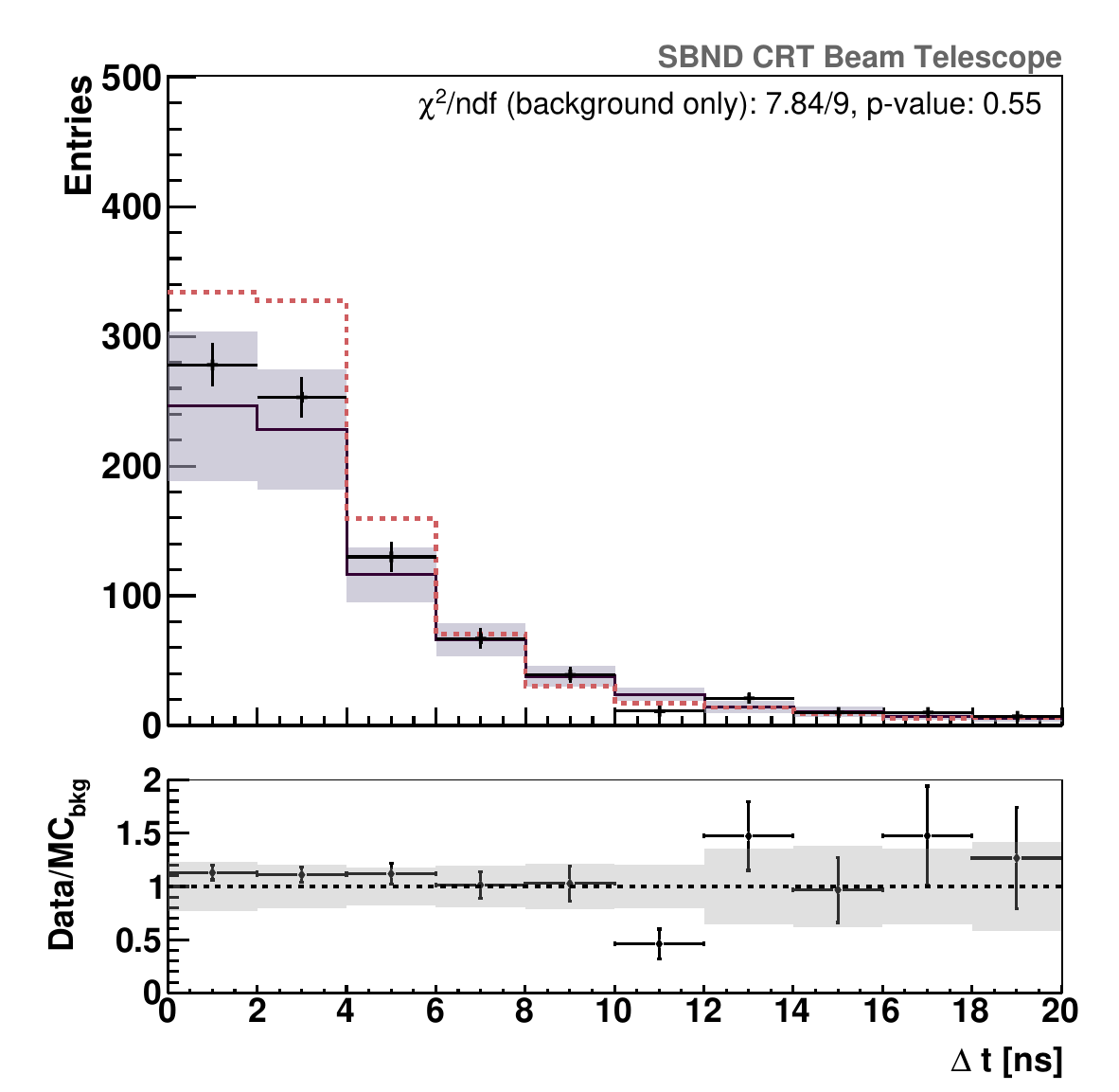}
        \label{fig:BDTVars_c}}
    \subfloat[]{
        \includegraphics[width=0.34\linewidth]{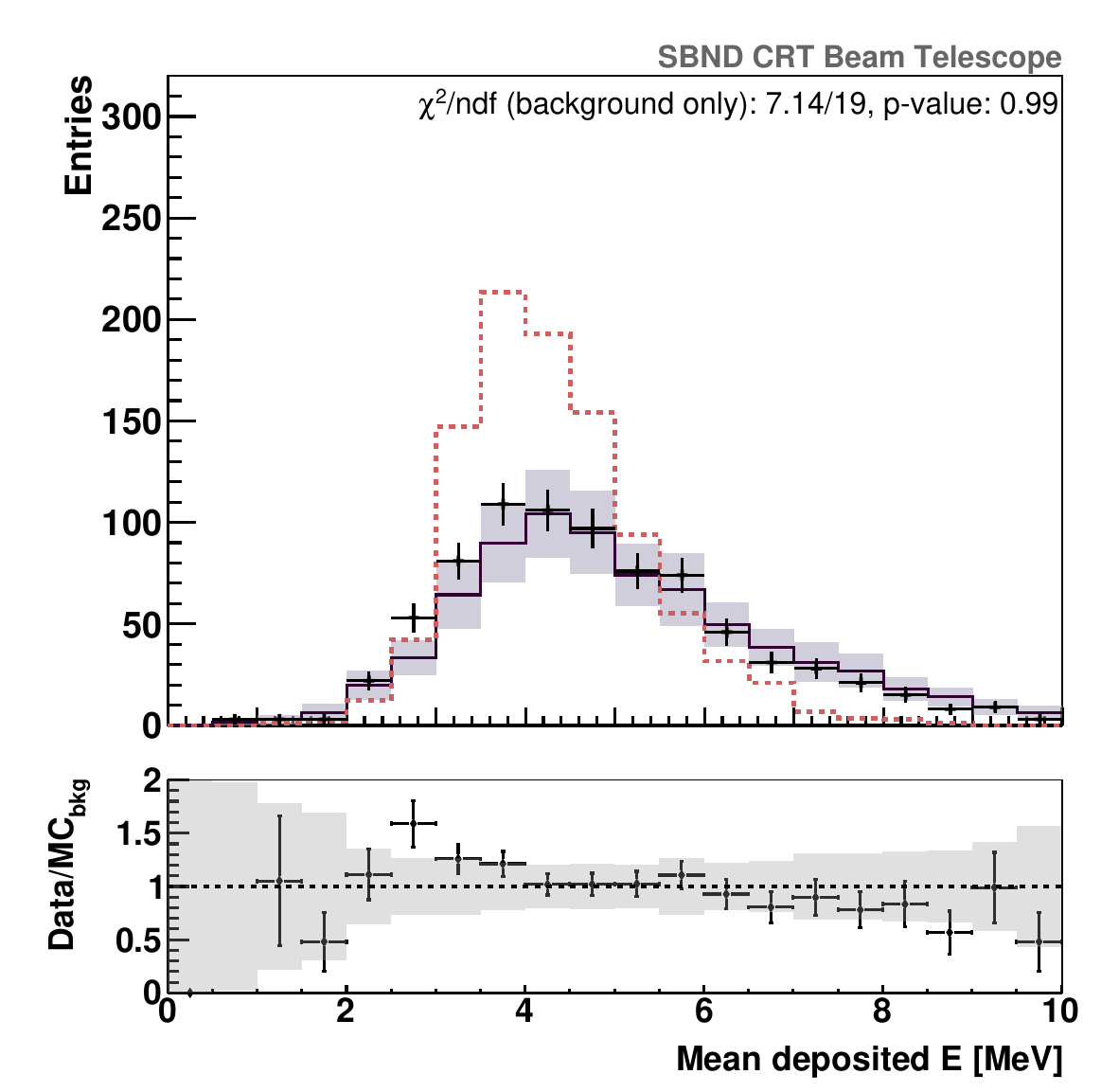}
        \label{fig:BDTVars_d}}
    \caption{Distributions of the four BDT input variables after minimal selection cuts: (a) three-dimensional distance, (b) horizontal (x-direction) separation, (c) timing difference $\Delta t$ between hits, and (d) mean deposited energy in each hit. The total neutrino-induced background prediction (black histogram with shaded uncertainty band) is compared to data (black points, $2.95\times10^{20}$ POT). The benchmark LLP signal ($m_N$ = 400 MeV, with decay length $c\tau=3.1~\textrm{m}$) is overlaid with a scaling factor of 50 for shape comparison. The ratio panels show the agreement between data and background. The uncertainty bands include flux, cross-section, and detector modeling systematics, and apply to the background only.}
    \label{fig:BDTVars}
\end{figure*}

Figure~\ref{fig:BDTVars} shows the distributions of these variables for a signal LLP with a mass of 400 MeV and a decay length of 3.1~m, as well as for the simulated neutrino background, after applying the minimal selections. Good agreement is observed between the background prediction and data for all variables, in both shape and normalization. Although no individual variable exhibits strong discriminating power on its own, the BDT exploits correlations among the input variables to capture higher-dimensional information, achieving the required separation. The resulting score distribution is then used directly in the final statistical analysis, detailed in~\cref{sec:results}.

%################################################
\section{Systematic Uncertainties}
\label{sec:systematics}
%################################################

Systematic uncertainties are evaluated for the simulated background prediction by comparing variations to the central value (CV) simulation sample. We consider the impact of neutrino flux simulation, neutrino interaction modeling, and detector modeling.

To assess the uncertainties on the neutrino flux prediction, the flux simulation from the MiniBooNE Collaboration is utilized~\cite{miniboone:2009flux, PhysRevLett.102.101802}, updated to the SBND CRT Beam Telescope location. For neutrino cross section modeling uncertainties, we use the event reweighting framework provided by \texttt{GENIE}~\cite{Andreopoulos:2015wxa}. For both cross section and flux systematics, we generate many MC replicas, called ``universes,'' where parameters in the models are varied within their uncertainties. Each universe represents a different reweighting. The simultaneous reweighting of all model parameters allows the correct treatment of correlations among them. We create $N$ such universes that can be combined to construct a covariance matrix:
\begin{equation}
\label{eq:covariance}
C_{ij} = \frac{1}{N} \sum_{n = 1}^{N} (b_i^n - b_i^\text{cv})(b_j^n - b_j^\text{cv}),
\end{equation}
where $b$ is a shorthand notation for the BDT score distribution, $i$ and $j$ are bin numbers, $b^\text{cv}_i$ is the predicted count in the CV BDT score bin $i$, and $b_i^n$ is the predicted count in the universe $n$ BDT score bin $i$. This method produces two covariance matrices, one for the neutrino flux and one for the neutrino interaction systematic. In the case of the flux, an additional 2\% uncertainty is added to account for POT counting uncertainty~\cite{miniboone:2009flux}. These matrices are then passed to the statistical analysis framework, described in \cref{sec:results}.

As detector model systematics are dominated by single detector parameters and are not possible to estimate through reweighting, a different procedure is used for this set of systematics uncertainties. The parameters used to simulate the CRT charge and timing response are derived from data and their uncertainties propagate to uncertainties in the detector modeling. To assess this source of systematic uncertainty, BDT score variations are generated by independently shifting each parameter by $\pm 1\sigma$, assuming no correlations, and running a new simulation that outputs BDT score distributions from new background sample variations generated with the new parameters. These distributions are then used as input to the statistical analysis framework, described in \cref{sec:results}.

The impact of flux and detector modeling systematic uncertainties on the signal are assessed in a dedicated limit-extraction study by assigning the signal the same fractional uncertainties as the background. As these uncertainties were found to have a negligible effect on the extracted limit, they are not included in the reported result.

\begin{table}
\caption{Contributions to the total systematic uncertainty for the simulated background sample, integrated over all BDT score bins. ``Sample statistics'' refers to the statistical uncertainty of the MC sample, normalized to the data POT. Entries are rounded to the nearest percent, while the total is evaluated from unrounded values summed in quadrature.}
\label{tab:all_syst}
    \begin{ruledtabular}
    \begin{tabular}{cc}
    Source of Uncertainty      & Relative Uncertainty  [\%]\\
    \hline
    Detector Modeling          &  16                    \\
    Neutrino Interaction       &  16                    \\
    Neutrino Flux              &  7                     \\
    Sample Statistics          &  5                     \\
    \hline
    Total                      &  25                    \\
    \end{tabular}
    \end{ruledtabular}
\end{table}
The flux, neutrino interaction, and detector systematic uncertainties on the background, integrated over all BDT score bins, are summarized in~\cref{tab:all_syst}, along with the contribution from each individual source.

%################################################
\section{Results and Discussion}
\label{sec:results}
%################################################

\begin{figure*}[!htbp]
    \centering
    \subfloat[]{
        \includegraphics[width=0.32\linewidth]{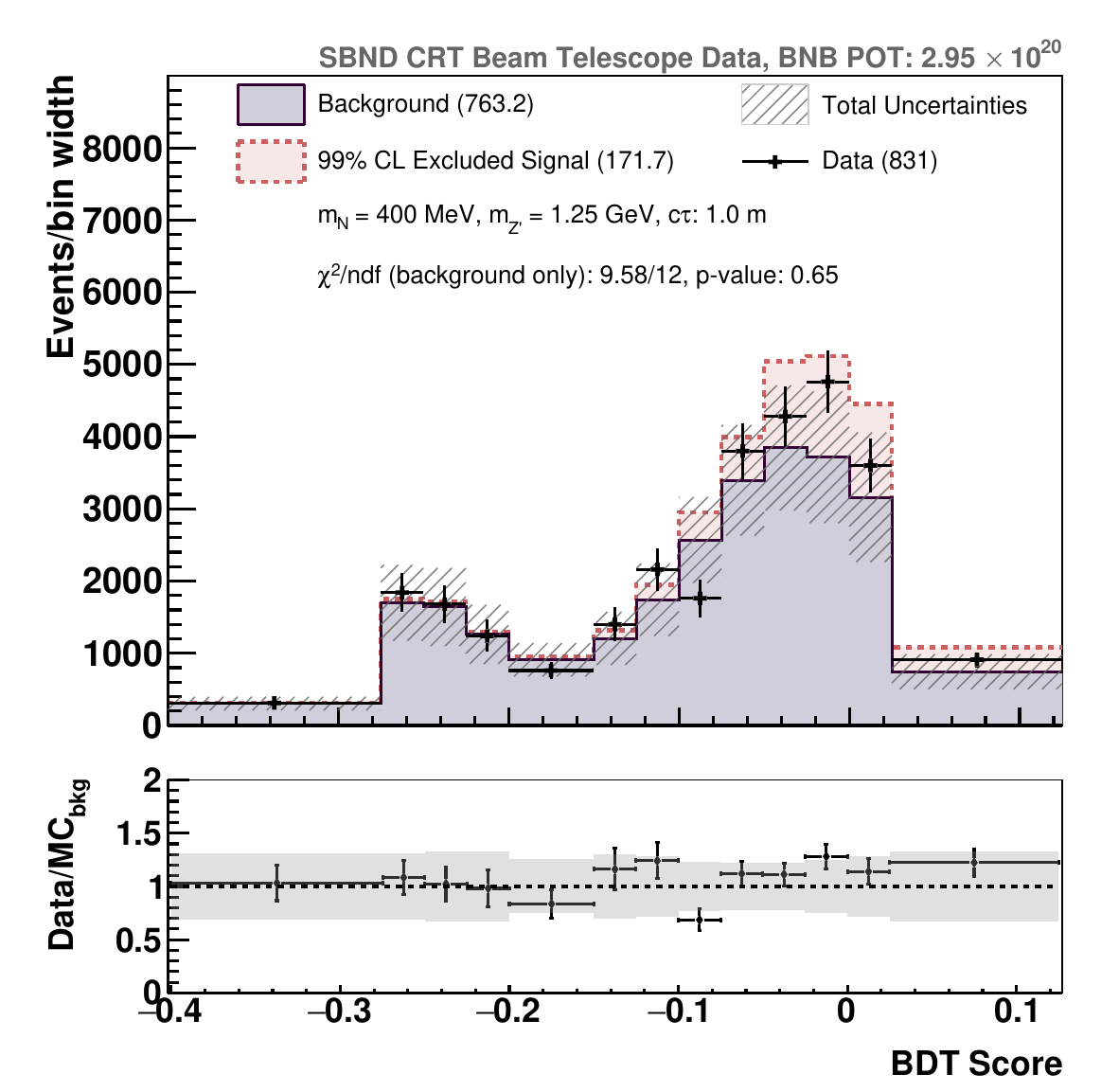}}
        \label{fig:FindBDT_a}
    \subfloat[]{
        \includegraphics[width=0.32\linewidth]{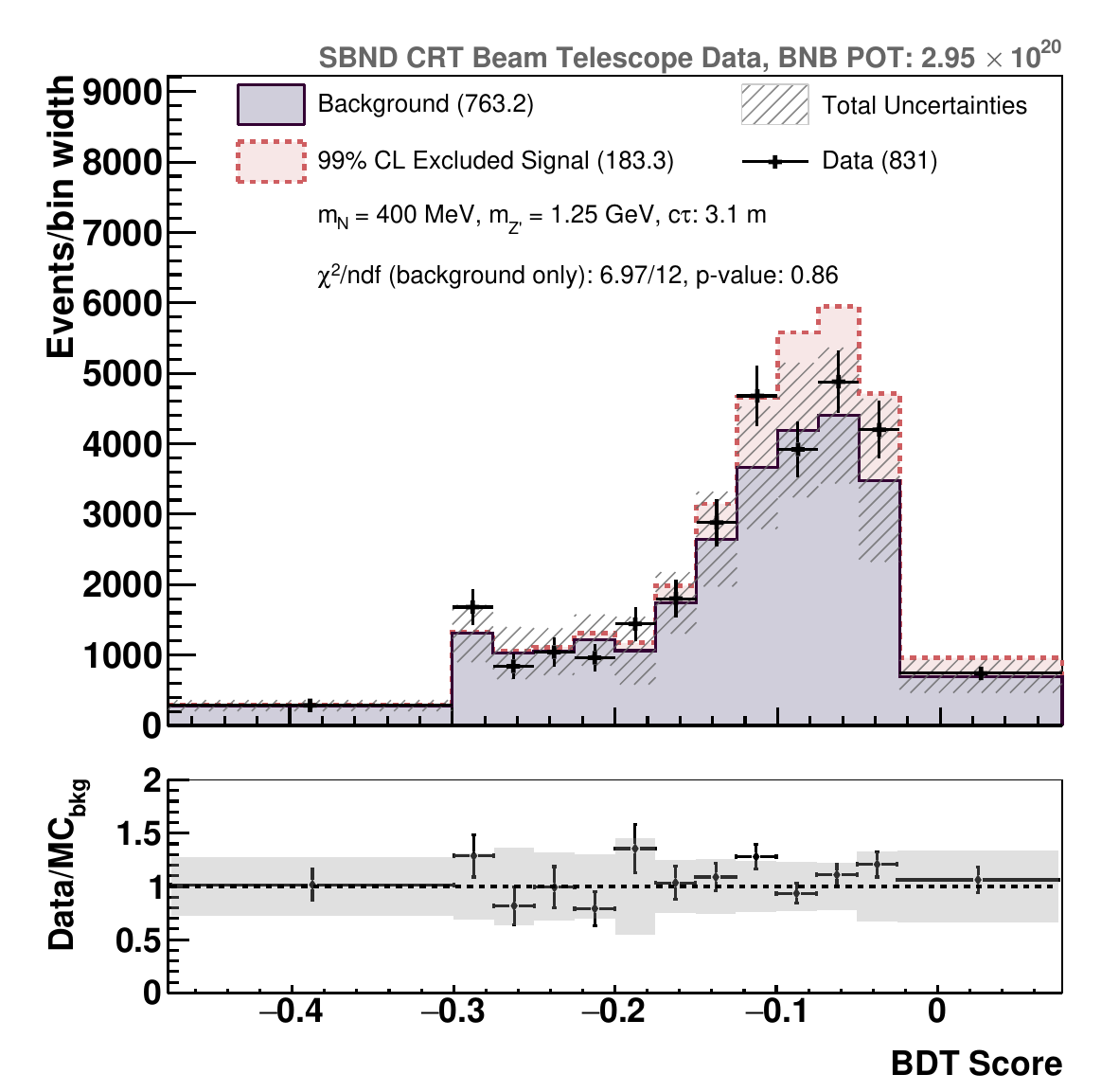}}
        \label{fig:FindBDT_b}
    \subfloat[]{
        \includegraphics[width=0.32\linewidth]{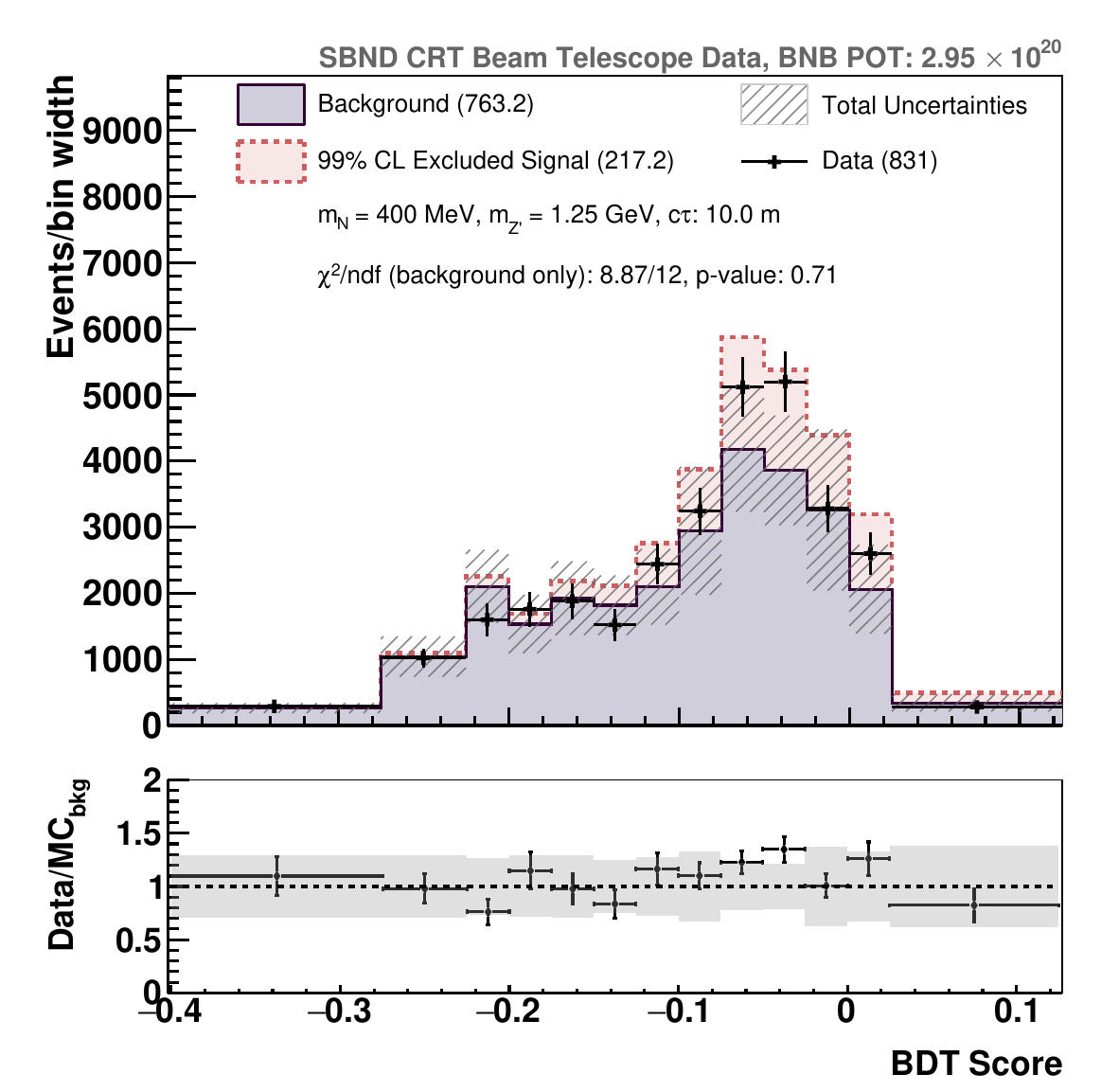}}
        \label{fig:FindBDT_c}
  \caption{BDT score distributions comparing data (black points, 831 events) to background-only predictions (gray shaded histogram, 763.2 events) for three representative LLP decay lengths: (a) $c\tau$=1~m, (b) $c\tau$=3.1~m, and (c) $c\tau$=10~m. The dotted red histogram shows the expected signal+background hypothesis, where the LLP signal is normalized to the 99\% $CL$ excluded event yield (171.7, 183.3, and 217.2 events respectively, as indicated in parentheses) and added to the background. Higher BDT scores (closer to zero) correspond to more signal-like events. The gray band represents total systematic uncertainties on the background prediction. The ratio panels show data and background ratio. Each BDT is trained specifically for the corresponding LLP decay length. Data correspond to $2.95\times10^{20}$ POT from the BNB.}
  \label{fig:FinalBDT}
\end{figure*}

The BDT score distributions for three representative LLP decay lengths after all selection criteria are shown in~\cref{fig:FinalBDT}. The red curves illustrate the signal+background hypothesis, where the expected LLP signal is normalized to the 99\% Confidence Level ($CL$) excluded yield and added to the background. The signal hypothesis predicts an enhancement of $\sim 171$ -- $217$ events (depending on decay length) relative to the 763.2 expected background events, which would manifest as a visible upturn in the rightmost bins of the distribution. The binning is chosen to balance statistical precision and discrimination of shape differences between signal and background, while avoiding bins with very low expected event counts. Alternative binning schemes were tested and found to yield consistent limits within statistical uncertainties. 

A small upward fluctuation is observed in the high-BDT-score region ($\text{BDT} > -0.2$), where LLP decays would preferentially accumulate. A goodness-of-fit test yields p-values of 0.65, 0.86, and 0.71 for decay lengths of 1~m, 3.1~m, and 10~m, confirming consistency with the background-only hypothesis. No statistically significant excess is observed, and the absence of such an excess allows us to set exclusion limits.

The statistical interpretation of the results follows a modified frequentist ($CL_s$) approach~\cite{JUNK1999435, ALRead_2002, Fisher:2006zz}, implemented using the \texttt{pyhf}~\cite{pyhf_joss_2021, cowan_asymptotic_2011} framework. At each LLP decay length, the \texttt{pyhf} package uses the BDT score distribution and scales the signal to set an upper limit on LLP production. 

The test statistic is defined using the log-likelihood ratio (LLR), which compares the signal-plus-background (S+B) hypothesis to the background-only (B) hypothesis. The p-values $CL_{S+B}$ and $CL_B$ are derived by integrating the LLR distribution under each hypothesis. A signal is excluded at 90\%  when ${CL_S = CL_{S+B}/CL_B < 0.1}$.

Systematic uncertainties on the background prediction are incorporated as nuisance parameters with Gaussian constraints. \texttt{pyhf} takes as input the deformations of the central-value spectrum and uses them to build template variations that parameterize the effect of each systematic uncertainty in the likelihood. For detector modeling, the variations described in \cref{sec:systematics} are directly incorporated into \texttt{pyhf}.

For neutrino flux and neutrino interaction model systematics, bin-to-bin correlations are instead encoded in the covariance matrix defined in \cref{eq:covariance}. Since \texttt{pyhf} only takes deformations of the spectrum, a principal component analysis (PCA) method~\cite{pearson1901pca, jolliffe2002pca} is applied, which decomposes the covariance matrix into a set of orthogonal eigenmodes. After PCA decomposition, only the most significant principal components, with eigenvalues larger than 0.5\% of the leading eigen-value, are retained as nuisance parameters in \texttt{pyhf}. These components are then passed to \texttt{pyhf} as shape variations of the CV spectrum.

The signal model used to determine exclusion limits assumes a specific neutrino upscattering cross section for LLP production and a specific branching ratio for the LLP decay into $e^+e^-$ with missing energy. The upper limit on the signal therefore constrains an effective cross section, derived from the product ${\sigma(\nu A \to N X) \times \text{Br}(N\to e^+e^-+E_{\mathrm{miss}})}$.

\begin{figure}[!htb]
    \centering
    \includegraphics[width=\linewidth]{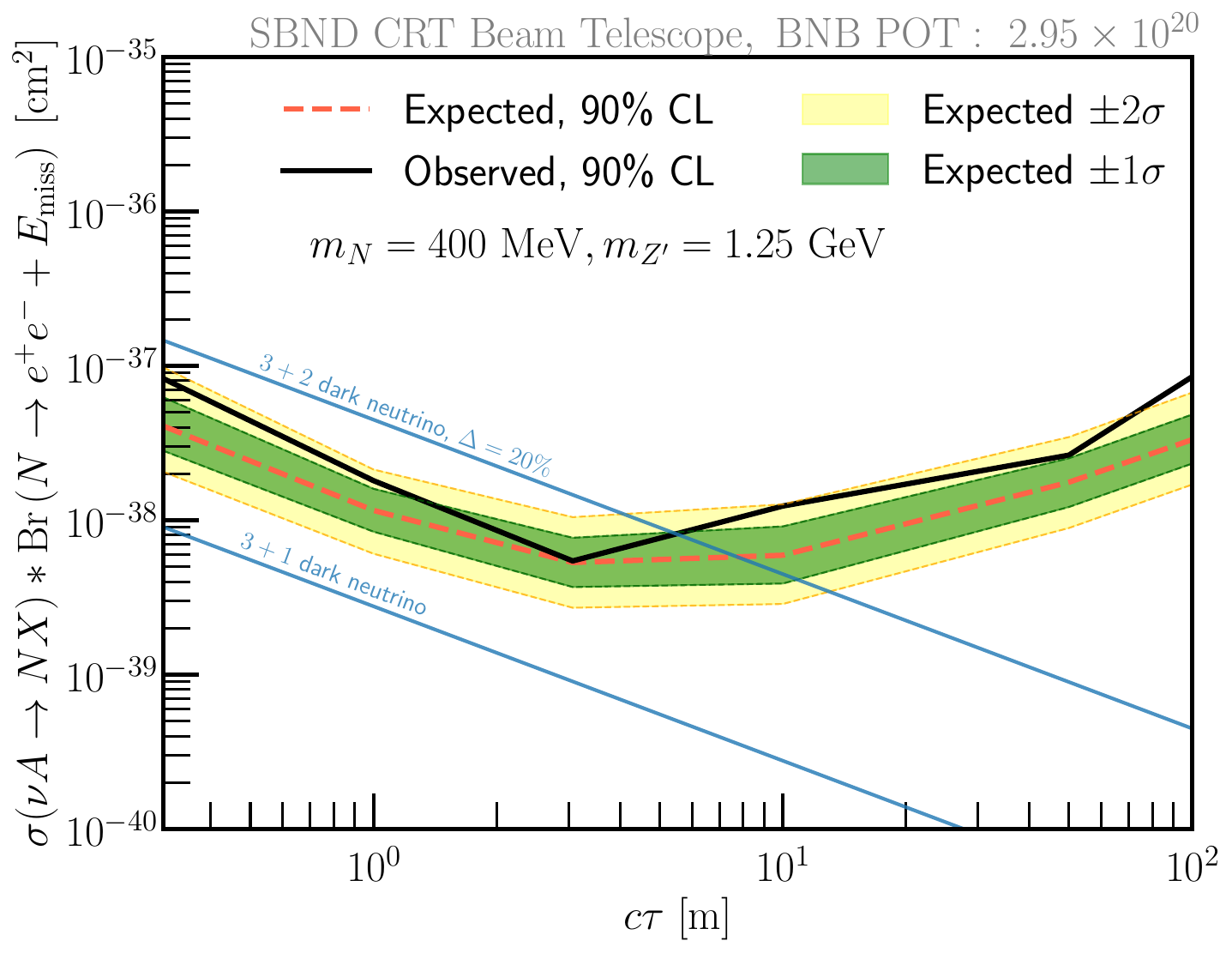}
    \caption{\label{fig:LLPResults} Upper limits (90\% $CL$) on $\sigma(\nu A \to NX)~\times~Br(N \to e^+e^- + E_{\mathrm{miss}})$ versus LLP decay length $c\tau$, showing observed (solid black) and expected (dashed red) exclusions with 1$\sigma$ (green) and 2$\sigma$ (yellow) bands from background fluctuations.
    The blue lines show representative dark-neutrino model trajectories for a minimal $3+1$ scenario and a $3+2$ scenario with relative heavy-neutrino mass splitting $\Delta\simeq20\%$, obtained by varying the kinetic-mixing parameter $\varepsilon$. Parameters for both scenarios are given in the text. 
    The limit is derived from the BDT score distribution, with the BDT trained separately for each $c\tau$ hypothesis but evaluated against a common background sample and the same observed dataset. The sensitivity is maximized for meter-scale lifetimes ($c\tau\sim$1–10~m), with reduced acceptance at shorter decay lengths (upstream decays) and longer decay lengths (geometric acceptance), as  discussed in \cref{sec:selection}. Results are based on $2.95\times 10^{20}$ POT from the BNB.}
\end{figure}

The observed and median expected 90\% $CL$ limits on the signal strength are shown versus LLP decay length in \cref{fig:LLPResults}.
We see that the SBND CRT Beam Telescope enables exclusions for LLPs with $\mathcal{O}(\mathrm{m})$ decay lengths -- this can be achieved in dark neutrino scenarios, especially ones with additional features such as other BSM states and/or interactions~\cite{Ballett:2019pyw}.  The resulting exclusions reach effective cross sections of order $\sim 5 \times 10^{-39}$~cm$^2$ at their strongest. To illustrate the interpretation of these limits in specific dark-neutrino scenarios~\cite{Bertuzzo:2018itn, Abdullahi:2023ejc}, \cref{fig:LLPResults} also shows representative model trajectories obtained by varying the kinetic-mixing parameter $\varepsilon$, for which the upscattering cross section scales as $\varepsilon^2$ while the dark neutrino lifetime scales approximately as $\varepsilon^{-2}$. The trajectories assume a dark-neutrino mass of 400~MeV, a dark gauge boson mass of 1.25~GeV, and a fixed mixing $|U_{\mu 4}|^2 = 4\times10^{-6}$ between the SM muon neutrino and the dark neutrino. The minimal $3+1$ scenario lies below the sensitivity of this search for this benchmark, while a representative $3+2$ scenario, in which the dark neutrino of mass $m_5$ decays into a lighter dark neutrino of mass $m_4$ through a coupling of the same strength, extends into the region probed by this analysis for a relative mass splitting $\Delta \equiv (m_5-m_4)/m_4 \simeq 20\%$. 

Several robustness checks were performed to assess the stability of the extracted limits. These include variations of the BDT score threshold, alternative binning configurations of the BDT distribution, and the exclusion of individual systematic uncertainty sources from the fit. In all cases, the resulting limits were found to be consistent within the expected statistical and systematic variations, demonstrating that the sensitivity is not dominated by a single analysis choice or systematic effect. As discussed above, the sensitivity is maximized for meter-scale decay lengths, reflecting the finite decay volume and geometric acceptance of the CRT Beam Telescope.

While these limits are weaker than those obtained in model-specific searches such as MicroBooNE~\cite{MicroBooNE:2025khi}, which probe both prompt and long-lived dark-neutrino decay regimes, the present search provides a complementary approach. As shown in~\cref{fig:LLPResults}, this search focuses on LLPs produced via neutrino upscattering in surrounding material and decaying at macroscopic distances, providing coverage of meter-scale decay lengths (0.3--100 m) using displaced decay signatures.

%################################################
\section{Conclusions}
\label{sec:conclusions}
%################################################

In this work, we present a search for LLPs using data collected with the SBND CRT Beam Telescope exposed to the BNB at Fermilab. The analysis targets LLPs produced via neutrino upscattering in surrounding material and decaying at macroscopic distances into visible final states, using timing, spatial and calorimetric information from the CRT system operating in a standalone configuration.

Using a data set corresponding to $2.95\times10^{20}$ POT, we set 90\% $CL$ upper limits on a generic class of LLPs produced via neutrino upscattering and decaying exclusively into an $e^+e^-+E_{\mathrm{miss}}$ final state. No statistically significant deviation from the background-only hypothesis is observed in the data. LLPs with laboratory decay lengths on the order of meters are excluded at the ${90\%~CL}$ for effective production cross sections at the level of $10^{-37}-10^{-39}$ cm$^2$, depending on the assumed decay length in the range 0.3--100~m.

This analysis constitutes the first physics beyond the Standard Model search performed with SBND data, and uses the subset of CRT modules as a standalone detector for LLP searches. Although the resulting exclusions are limited by the geometric acceptance and the event information available in the CRT-only configuration, the close proximity of SBND to the BNB target enables meaningful sensitivity, demonstrating the feasibility of performing LLP searches with detector subsystems not originally designed for this purpose.

Building on the present analysis, future studies using data from the full SBND detector and combining information from the liquid argon TPC, CRT, and photon detection systems could extend these exclusions to a broader range of LLP decay lengths and topologies.

%################################################
\begin{acknowledgments}
This document was prepared by the SBND Collaboration using the resources of the Fermi National Accelerator Laboratory (Fermilab), a U.S. Department of Energy, Office of Science, Office of High Energy Physics HEP User Facility. Fermilab is managed by FermiForward Discovery Group, LLC, acting under Contract No. 89243024CSC000002. The SBND Collaboration acknowledges the generous support of the following organizations: the U.S. Department of Energy, Office of Science, Office of High Energy Physics; the U.S. National Science Foundation; Los Alamos National Laboratory for LDRD funding, the Science and Technology Facilities Council (STFC), part of United Kingdom Research and Innovation (UKRI), the UKRI Future Leaders Fellowship (grant number MR/V022407/1), and The Royal Society; the Swiss National Science Foundation; the Spanish Ministerio de Ciencia, Innovacíon y Universidades (MICIU/ AEI/ 10.13039/ 501100011033) under grants No PRE2019-090468, CNS2022-136022, RYC2022-036471-I, PID2023-147949NB-C51 \& C53 and Comunidad de Madrid (PEJ-2023-AI/COM-28399); the European Union’s Horizon 2020 research and innovation program under GA no 101004761 and the Marie Sklodowska-Curie grant agreements No 822185, 101081478, and 101003460; the São Paulo Research Foundation 1098 (FAPESP), the National Council of Scientific and Technological Development (CNPq) and Ministry of Science, Technology \& Innovations-MCTI of Brazil; the Minas Gerais research foundation (FAPEMIG), grants APQ-00544-23 and APQ-01249-24; the Anusandhan National Research Foundation (ANRF, India) under the Ramanujan Fellowship (Grant No. RJF/2025/000203). An award of computer time was provided by the ASCR Leadership Computing Challenge (ALCC) program. This research used resources of the Argonne Leadership Computing Facility, which is a U.S. Department of Energy Office of Science User Facility operated under contract DE-AC02-06CH11357.
\end{acknowledgments}
%################################################

%################################################
\bibliography{main}

\begin{thebibliography}{36}%
\makeatletter
\providecommand \@ifxundefined [1]{%
 \@ifx{#1\undefined}
}%
\providecommand \@ifnum [1]{%
 \ifnum #1\expandafter \@firstoftwo
 \else \expandafter \@secondoftwo
 \fi
}%
\providecommand \@ifx [1]{%
 \ifx #1\expandafter \@firstoftwo
 \else \expandafter \@secondoftwo
 \fi
}%
\providecommand \natexlab [1]{#1}%
\providecommand \enquote  [1]{``#1''}%
\providecommand \bibnamefont  [1]{#1}%
\providecommand \bibfnamefont [1]{#1}%
\providecommand \citenamefont [1]{#1}%
\providecommand \href@noop [0]{\@secondoftwo}%
\providecommand \href [0]{\begingroup \@sanitize@url \@href}%
\providecommand \@href[1]{\@@startlink{#1}\@@href}%
\providecommand \@@href[1]{\endgroup#1\@@endlink}%
\providecommand \@sanitize@url [0]{\catcode `\\12\catcode `\$12\catcode `\&12\catcode `\#12\catcode `\^12\catcode `\_12\catcode `\%12\relax}%
\providecommand \@@startlink[1]{}%
\providecommand \@@endlink[0]{}%
\providecommand \url  [0]{\begingroup\@sanitize@url \@url }%
\providecommand \@url [1]{\endgroup\@href {#1}{\urlprefix }}%
\providecommand \urlprefix  [0]{URL }%
\providecommand \Eprint [0]{\href }%
\providecommand \doibase [0]{https://doi.org/}%
\providecommand \selectlanguage [0]{\@gobble}%
\providecommand \bibinfo  [0]{\@secondoftwo}%
\providecommand \bibfield  [0]{\@secondoftwo}%
\providecommand \translation [1]{[#1]}%
\providecommand \BibitemOpen [0]{}%
\providecommand \bibitemStop [0]{}%
\providecommand \bibitemNoStop [0]{.\EOS\space}%
\providecommand \EOS [0]{\spacefactor3000\relax}%
\providecommand \BibitemShut  [1]{\csname bibitem#1\endcsname}%
\let\auto@bib@innerbib\@empty
%</preamble>
\bibitem [{\citenamefont {Abratenko}\ \emph {et~al.}(2026{\natexlab{a}})\citenamefont {Abratenko} \emph {et~al.}}]{sbnd:2026detector}%
  \BibitemOpen
  \bibfield  {author} {\bibinfo {author} {\bibfnamefont {P.}~\bibnamefont {Abratenko}} \emph {et~al.} (\bibinfo {collaboration} {SBND}),\ }\href@noop {} {\bibinfo {title} {{Short-Baseline Near Detector (SBND): Design and Initial Performance}}} (\bibinfo {year} {2026}{\natexlab{a}}),\ \Eprint {https://arxiv.org/abs/2610.11782} {arXiv:2610.11782 [hep-ex]} \BibitemShut {NoStop}%
\bibitem [{\citenamefont {Aguilar-Arevalo}\ \emph {et~al.}(2009{\natexlab{a}})\citenamefont {Aguilar-Arevalo} \emph {et~al.}}]{miniboone:2009flux}%
  \BibitemOpen
  \bibfield  {author} {\bibinfo {author} {\bibfnamefont {A.~A.}\ \bibnamefont {Aguilar-Arevalo}} \emph {et~al.} (\bibinfo {collaboration} {{MiniBooNE}}),\ }\bibfield  {title} {\bibinfo {title} {{Neutrino flux prediction at MiniBooNE}},\ }\href {https://doi.org/10.1103/PhysRevD.79.072002} {\bibfield  {journal} {\bibinfo  {journal} {Phys. Rev. D}\ }\textbf {\bibinfo {volume} {79}},\ \bibinfo {pages} {072002} (\bibinfo {year} {2009}{\natexlab{a}})},\ \Eprint {https://arxiv.org/abs/0806.1449} {arXiv:0806.1449 [hep-ex]} \BibitemShut {NoStop}%
\bibitem [{\citenamefont {Acciarri}\ \emph {et~al.}(2015)\citenamefont {Acciarri} \emph {et~al.}}]{MicroBooNE:2015bmn}%
  \BibitemOpen
  \bibfield  {author} {\bibinfo {author} {\bibfnamefont {R.}~\bibnamefont {Acciarri}} \emph {et~al.} (\bibinfo {collaboration} {{MicroBooNE, LAr1-ND, ICARUS-WA104}}),\ }\href@noop {} {\bibinfo {title} {{A Proposal for a Three Detector Short-Baseline Neutrino Oscillation Program in the Fermilab Booster Neutrino Beam}}} (\bibinfo {year} {2015}),\ \Eprint {https://arxiv.org/abs/1503.01520} {arXiv:1503.01520 [physics.ins-det]} \BibitemShut {NoStop}%
\bibitem [{\citenamefont {Machado}\ \emph {et~al.}(2019)\citenamefont {Machado}, \citenamefont {Palamara},\ and\ \citenamefont {Schmitz}}]{Machado:2019oxb}%
  \BibitemOpen
  \bibfield  {author} {\bibinfo {author} {\bibfnamefont {P.~A.}\ \bibnamefont {Machado}}, \bibinfo {author} {\bibfnamefont {O.}~\bibnamefont {Palamara}},\ and\ \bibinfo {author} {\bibfnamefont {D.~W.}\ \bibnamefont {Schmitz}},\ }\bibfield  {title} {\bibinfo {title} {{The Short-Baseline Neutrino Program at Fermilab}},\ }\href {https://doi.org/10.1146/annurev-nucl-101917-020949} {\bibfield  {journal} {\bibinfo  {journal} {Ann. Rev. Nucl. Part. Sci.}\ }\textbf {\bibinfo {volume} {69}},\ \bibinfo {pages} {363} (\bibinfo {year} {2019})},\ \Eprint {https://arxiv.org/abs/1903.04608} {arXiv:1903.04608 [hep-ex]} \BibitemShut {NoStop}%
\bibitem [{\citenamefont {Abratenko}\ \emph {et~al.}(2024{\natexlab{a}})\citenamefont {Abratenko} \emph {et~al.}}]{MicroBooNE:2023eef}%
  \BibitemOpen
  \bibfield  {author} {\bibinfo {author} {\bibfnamefont {P.}~\bibnamefont {Abratenko}} \emph {et~al.} (\bibinfo {collaboration} {MicroBooNE}),\ }\bibfield  {title} {\bibinfo {title} {{Search for Heavy Neutral Leptons in Electron-Positron and Neutral-Pion Final States with the MicroBooNE Detector}},\ }\href {https://doi.org/10.1103/PhysRevLett.132.041801} {\bibfield  {journal} {\bibinfo  {journal} {Phys. Rev. Lett.}\ }\textbf {\bibinfo {volume} {132}},\ \bibinfo {pages} {041801} (\bibinfo {year} {2024}{\natexlab{a}})},\ \Eprint {https://arxiv.org/abs/2310.07660} {arXiv:2310.07660 [hep-ex]} \BibitemShut {NoStop}%
\bibitem [{\citenamefont {Abratenko}\ \emph {et~al.}(2022{\natexlab{a}})\citenamefont {Abratenko} \emph {et~al.}}]{MicroBooNE:2022ctm}%
  \BibitemOpen
  \bibfield  {author} {\bibinfo {author} {\bibfnamefont {P.}~\bibnamefont {Abratenko}} \emph {et~al.} (\bibinfo {collaboration} {MicroBooNE}),\ }\bibfield  {title} {\bibinfo {title} {{Search for long-lived heavy neutral leptons and Higgs portal scalars decaying in the MicroBooNE detector}},\ }\href {https://doi.org/10.1103/PhysRevD.106.092006} {\bibfield  {journal} {\bibinfo  {journal} {Phys. Rev. D}\ }\textbf {\bibinfo {volume} {106}},\ \bibinfo {pages} {092006} (\bibinfo {year} {2022}{\natexlab{a}})},\ \Eprint {https://arxiv.org/abs/2207.03840} {arXiv:2207.03840 [hep-ex]} \BibitemShut {NoStop}%
\bibitem [{\citenamefont {Abratenko}\ \emph {et~al.}(2024{\natexlab{b}})\citenamefont {Abratenko} \emph {et~al.}}]{MicroBooNE:2023gmv}%
  \BibitemOpen
  \bibfield  {author} {\bibinfo {author} {\bibfnamefont {P.}~\bibnamefont {Abratenko}} \emph {et~al.} (\bibinfo {collaboration} {MicroBooNE}),\ }\bibfield  {title} {\bibinfo {title} {{First Search for Dark-Trident Processes Using the MicroBooNE Detector}},\ }\href {https://doi.org/10.1103/PhysRevLett.132.241801} {\bibfield  {journal} {\bibinfo  {journal} {Phys. Rev. Lett.}\ }\textbf {\bibinfo {volume} {132}},\ \bibinfo {pages} {241801} (\bibinfo {year} {2024}{\natexlab{b}})},\ \Eprint {https://arxiv.org/abs/2312.13945} {arXiv:2312.13945 [hep-ex]} \BibitemShut {NoStop}%
\bibitem [{\citenamefont {Abe}\ \emph {et~al.}(2019)\citenamefont {Abe} \emph {et~al.}}]{T2K:2019jwa}%
  \BibitemOpen
  \bibfield  {author} {\bibinfo {author} {\bibfnamefont {K.}~\bibnamefont {Abe}} \emph {et~al.} (\bibinfo {collaboration} {T2K}),\ }\bibfield  {title} {\bibinfo {title} {{Search for heavy neutrinos with the T2K near detector ND280}},\ }\href {https://doi.org/10.1103/PhysRevD.100.052006} {\bibfield  {journal} {\bibinfo  {journal} {Phys. Rev. D}\ }\textbf {\bibinfo {volume} {100}},\ \bibinfo {pages} {052006} (\bibinfo {year} {2019})},\ \Eprint {https://arxiv.org/abs/1902.07598} {arXiv:1902.07598 [hep-ex]} \BibitemShut {NoStop}%
\bibitem [{\citenamefont {Alrahman}\ \emph {et~al.}(2025)\citenamefont {Alrahman} \emph {et~al.}}]{ICARUS:2024oqb}%
  \BibitemOpen
  \bibfield  {author} {\bibinfo {author} {\bibfnamefont {F.~A.}\ \bibnamefont {Alrahman}} \emph {et~al.} (\bibinfo {collaboration} {ICARUS}),\ }\bibfield  {title} {\bibinfo {title} {{Search for a Hidden Sector Scalar from Kaon Decay in the Dimuon Final State at ICARUS}},\ }\href {https://doi.org/10.1103/PhysRevLett.134.151801} {\bibfield  {journal} {\bibinfo  {journal} {Phys. Rev. Lett.}\ }\textbf {\bibinfo {volume} {134}},\ \bibinfo {pages} {151801} (\bibinfo {year} {2025})},\ \Eprint {https://arxiv.org/abs/2411.02727} {arXiv:2411.02727 [hep-ex]} \BibitemShut {NoStop}%
\bibitem [{\citenamefont {Bertuzzo}\ \emph {et~al.}(2018)\citenamefont {Bertuzzo}, \citenamefont {Jana}, \citenamefont {Machado},\ and\ \citenamefont {Zukanovich~Funchal}}]{Bertuzzo:2018itn}%
  \BibitemOpen
  \bibfield  {author} {\bibinfo {author} {\bibfnamefont {E.}~\bibnamefont {Bertuzzo}}, \bibinfo {author} {\bibfnamefont {S.}~\bibnamefont {Jana}}, \bibinfo {author} {\bibfnamefont {P.~A.~N.}\ \bibnamefont {Machado}},\ and\ \bibinfo {author} {\bibfnamefont {R.}~\bibnamefont {Zukanovich~Funchal}},\ }\bibfield  {title} {\bibinfo {title} {{Dark Neutrino Portal to Explain MiniBooNE excess}},\ }\href {https://doi.org/10.1103/PhysRevLett.121.241801} {\bibfield  {journal} {\bibinfo  {journal} {Phys. Rev. Lett.}\ }\textbf {\bibinfo {volume} {121}},\ \bibinfo {pages} {241801} (\bibinfo {year} {2018})},\ \Eprint {https://arxiv.org/abs/1807.09877} {arXiv:1807.09877 [hep-ph]} \BibitemShut {NoStop}%
\bibitem [{\citenamefont {Ballett}\ \emph {et~al.}(2020)\citenamefont {Ballett}, \citenamefont {Hostert},\ and\ \citenamefont {Pascoli}}]{Ballett:2019pyw}%
  \BibitemOpen
  \bibfield  {author} {\bibinfo {author} {\bibfnamefont {P.}~\bibnamefont {Ballett}}, \bibinfo {author} {\bibfnamefont {M.}~\bibnamefont {Hostert}},\ and\ \bibinfo {author} {\bibfnamefont {S.}~\bibnamefont {Pascoli}},\ }\bibfield  {title} {\bibinfo {title} {{Dark Neutrinos and a Three Portal Connection to the Standard Model}},\ }\href {https://doi.org/10.1103/PhysRevD.101.115025} {\bibfield  {journal} {\bibinfo  {journal} {Phys. Rev. D}\ }\textbf {\bibinfo {volume} {101}},\ \bibinfo {pages} {115025} (\bibinfo {year} {2020})},\ \Eprint {https://arxiv.org/abs/1903.07589} {arXiv:1903.07589 [hep-ph]} \BibitemShut {NoStop}%
\bibitem [{\citenamefont {Auger}\ \emph {et~al.}(2017)\citenamefont {Auger} \emph {et~al.}}]{Auger:2016tjc}%
  \BibitemOpen
  \bibfield  {author} {\bibinfo {author} {\bibfnamefont {M.}~\bibnamefont {Auger}} \emph {et~al.},\ }\bibfield  {title} {\bibinfo {title} {{A Novel Cosmic Ray Tagger System for Liquid Argon TPC Neutrino Detectors}},\ }\href {https://doi.org/10.3390/instruments1010002} {\bibfield  {journal} {\bibinfo  {journal} {Instruments}\ }\textbf {\bibinfo {volume} {1}},\ \bibinfo {pages} {2} (\bibinfo {year} {2017})},\ \Eprint {https://arxiv.org/abs/1612.04614} {arXiv:1612.04614 [physics.ins-det]} \BibitemShut {NoStop}%
\bibitem [{Ham()}]{Hamamatsu}%
  \BibitemOpen
  \href@noop {} {\bibinfo {title} {{Hamamatsu S12825-050P}}},\ \bibinfo {howpublished} {\url{ http://www.hamamatsu.com/jp/en/S12571-050P.html}},\ \bibinfo {note} {accessed: 2023-09-13}\BibitemShut {NoStop}%
\bibitem [{\citenamefont {Auger}\ \emph {et~al.}(2016)\citenamefont {Auger}, \citenamefont {Ereditato}, \citenamefont {Goeldi}, \citenamefont {Kreslo}, \citenamefont {Lorca}, \citenamefont {Luethi}, \citenamefont {von Rohr}, \citenamefont {Sinclair},\ and\ \citenamefont {Weber}}]{Auger:2016vpo}%
  \BibitemOpen
  \bibfield  {author} {\bibinfo {author} {\bibfnamefont {M.}~\bibnamefont {Auger}}, \bibinfo {author} {\bibfnamefont {A.}~\bibnamefont {Ereditato}}, \bibinfo {author} {\bibfnamefont {D.}~\bibnamefont {Goeldi}}, \bibinfo {author} {\bibfnamefont {I.}~\bibnamefont {Kreslo}}, \bibinfo {author} {\bibfnamefont {D.}~\bibnamefont {Lorca}}, \bibinfo {author} {\bibfnamefont {M.}~\bibnamefont {Luethi}}, \bibinfo {author} {\bibfnamefont {C.~R.}\ \bibnamefont {von Rohr}}, \bibinfo {author} {\bibfnamefont {J.}~\bibnamefont {Sinclair}},\ and\ \bibinfo {author} {\bibfnamefont {M.~S.}\ \bibnamefont {Weber}},\ }\bibfield  {title} {\bibinfo {title} {{Multi-channel front-end board for SiPM readout}},\ }\href {https://doi.org/10.1088/1748-0221/11/10/P10005} {\bibfield  {journal} {\bibinfo  {journal} {JINST}\ }\textbf {\bibinfo {volume} {11}}\bibfield  {number} {\bibinfo  {number} { (10)},\ \bibinfo {pages} {P10005}},\ }\Eprint {https://arxiv.org/abs/1606.02290} {arXiv:1606.02290 [physics.ins-det]} \BibitemShut {NoStop}%
\bibitem [{\citenamefont {Abratenko}\ \emph {et~al.}(2026{\natexlab{b}})\citenamefont {Abratenko} \emph {et~al.}}]{SBND:2025plf}%
  \BibitemOpen
  \bibfield  {author} {\bibinfo {author} {\bibfnamefont {P.}~\bibnamefont {Abratenko}} \emph {et~al.} (\bibinfo {collaboration} {SBND}),\ }\bibfield  {title} {\bibinfo {title} {{Sampling off-axis neutrino fluxes with the short-baseline near detector}},\ }\href {https://doi.org/10.1103/hc3v-tqdb} {\bibfield  {journal} {\bibinfo  {journal} {Phys. Rev. D}\ }\textbf {\bibinfo {volume} {113}},\ \bibinfo {pages} {072007} (\bibinfo {year} {2026}{\natexlab{b}})}\BibitemShut {NoStop}%
\bibitem [{\citenamefont {Ballett}\ \emph {et~al.}(2019)\citenamefont {Ballett}, \citenamefont {Pascoli},\ and\ \citenamefont {Ross-Lonergan}}]{Ballett:2018ynz}%
  \BibitemOpen
  \bibfield  {author} {\bibinfo {author} {\bibfnamefont {P.}~\bibnamefont {Ballett}}, \bibinfo {author} {\bibfnamefont {S.}~\bibnamefont {Pascoli}},\ and\ \bibinfo {author} {\bibfnamefont {M.}~\bibnamefont {Ross-Lonergan}},\ }\bibfield  {title} {\bibinfo {title} {{U(1)' mediated decays of heavy sterile neutrinos in MiniBooNE}},\ }\href {https://doi.org/10.1103/PhysRevD.99.071701} {\bibfield  {journal} {\bibinfo  {journal} {Phys. Rev. D}\ }\textbf {\bibinfo {volume} {99}},\ \bibinfo {pages} {071701} (\bibinfo {year} {2019})},\ \Eprint {https://arxiv.org/abs/1808.02915} {arXiv:1808.02915 [hep-ph]} \BibitemShut {NoStop}%
\bibitem [{\citenamefont {Abdullahi}\ \emph {et~al.}(2021)\citenamefont {Abdullahi}, \citenamefont {Hostert},\ and\ \citenamefont {Pascoli}}]{Abdullahi:2020nyr}%
  \BibitemOpen
  \bibfield  {author} {\bibinfo {author} {\bibfnamefont {A.}~\bibnamefont {Abdullahi}}, \bibinfo {author} {\bibfnamefont {M.}~\bibnamefont {Hostert}},\ and\ \bibinfo {author} {\bibfnamefont {S.}~\bibnamefont {Pascoli}},\ }\bibfield  {title} {\bibinfo {title} {{A dark seesaw solution to low energy anomalies: MiniBooNE, the muon (g {\ensuremath{-}} 2), and BaBar}},\ }\href {https://doi.org/10.1016/j.physletb.2021.136531} {\bibfield  {journal} {\bibinfo  {journal} {Phys. Lett. B}\ }\textbf {\bibinfo {volume} {820}},\ \bibinfo {pages} {136531} (\bibinfo {year} {2021})},\ \Eprint {https://arxiv.org/abs/2007.11813} {arXiv:2007.11813 [hep-ph]} \BibitemShut {NoStop}%
\bibitem [{\citenamefont {Dutta}\ \emph {et~al.}(2025)\citenamefont {Dutta}, \citenamefont {Goswami}, \citenamefont {Karthikeyan},\ and\ \citenamefont {Kelly}}]{Dutta:2025npn}%
  \BibitemOpen
  \bibfield  {author} {\bibinfo {author} {\bibfnamefont {B.}~\bibnamefont {Dutta}}, \bibinfo {author} {\bibfnamefont {D.}~\bibnamefont {Goswami}}, \bibinfo {author} {\bibfnamefont {A.}~\bibnamefont {Karthikeyan}},\ and\ \bibinfo {author} {\bibfnamefont {K.~J.}\ \bibnamefont {Kelly}},\ }\bibfield  {title} {\bibinfo {title} {{Dirt/Detector/Dump: complementary BSM production at Short-Baseline Neutrino Facilities}},\ }\href {https://doi.org/10.1007/JHEP05(2025)240} {\bibfield  {journal} {\bibinfo  {journal} {JHEP}\ }\textbf {\bibinfo {volume} {05}},\ \bibinfo {pages} {240}},\ \Eprint {https://arxiv.org/abs/2501.09840} {arXiv:2501.09840 [hep-ph]} \BibitemShut {NoStop}%
\bibitem [{\citenamefont {Abdullahi}\ \emph {et~al.}(2024)\citenamefont {Abdullahi}, \citenamefont {Hoefken~Zink}, \citenamefont {Hostert}, \citenamefont {Massaro},\ and\ \citenamefont {Pascoli}}]{Abdullahi:2022cdw}%
  \BibitemOpen
  \bibfield  {author} {\bibinfo {author} {\bibfnamefont {A.~M.}\ \bibnamefont {Abdullahi}}, \bibinfo {author} {\bibfnamefont {J.}~\bibnamefont {Hoefken~Zink}}, \bibinfo {author} {\bibfnamefont {M.}~\bibnamefont {Hostert}}, \bibinfo {author} {\bibfnamefont {D.}~\bibnamefont {Massaro}},\ and\ \bibinfo {author} {\bibfnamefont {S.}~\bibnamefont {Pascoli}},\ }\bibfield  {title} {\bibinfo {title} {{DarkNews: A Python-based event generator for heavy neutral lepton production in neutrino-nucleus scattering}},\ }\href {https://doi.org/10.1016/j.cpc.2023.109075} {\bibfield  {journal} {\bibinfo  {journal} {Comput. Phys. Commun.}\ }\textbf {\bibinfo {volume} {297}},\ \bibinfo {pages} {109075} (\bibinfo {year} {2024})},\ \Eprint {https://arxiv.org/abs/2207.04137} {arXiv:2207.04137 [hep-ph]} \BibitemShut {NoStop}%
\bibitem [{\citenamefont {Andreopoulos}\ \emph {et~al.}(2010)\citenamefont {Andreopoulos} \emph {et~al.}}]{Andreopoulos:2009rq}%
  \BibitemOpen
  \bibfield  {author} {\bibinfo {author} {\bibfnamefont {C.}~\bibnamefont {Andreopoulos}} \emph {et~al.},\ }\bibfield  {title} {\bibinfo {title} {{The GENIE Neutrino Monte Carlo Generator}},\ }\href {https://doi.org/10.1016/j.nima.2009.12.009} {\bibfield  {journal} {\bibinfo  {journal} {Nucl. Instrum. Meth. A}\ }\textbf {\bibinfo {volume} {614}},\ \bibinfo {pages} {87} (\bibinfo {year} {2010})},\ \Eprint {https://arxiv.org/abs/0905.2517} {arXiv:0905.2517 [hep-ph]} \BibitemShut {NoStop}%
\bibitem [{\citenamefont {Abratenko}\ \emph {et~al.}(2022{\natexlab{b}})\citenamefont {Abratenko} \emph {et~al.}}]{PhysRevD.105.072001}%
  \BibitemOpen
  \bibfield  {author} {\bibinfo {author} {\bibfnamefont {P.}~\bibnamefont {Abratenko}} \emph {et~al.} (\bibinfo {collaboration} {{MicroBooNE}}),\ }\bibfield  {title} {\bibinfo {title} {{New $\mathrm{CC}0\ensuremath{\pi}$ GENIE model tune for MicroBooNE}},\ }\href {https://doi.org/10.1103/PhysRevD.105.072001} {\bibfield  {journal} {\bibinfo  {journal} {Phys. Rev. D}\ }\textbf {\bibinfo {volume} {105}},\ \bibinfo {pages} {072001} (\bibinfo {year} {2022}{\natexlab{b}})}\BibitemShut {NoStop}%
\bibitem [{\citenamefont {Agostinelli}\ \emph {et~al.}(2003)\citenamefont {Agostinelli} \emph {et~al.}}]{GEANT4:2002zbu}%
  \BibitemOpen
  \bibfield  {author} {\bibinfo {author} {\bibfnamefont {S.}~\bibnamefont {Agostinelli}} \emph {et~al.} (\bibinfo {collaboration} {{GEANT4}}),\ }\bibfield  {title} {\bibinfo {title} {{GEANT4--a simulation toolkit}},\ }\href {https://doi.org/10.1016/S0168-9002(03)01368-8} {\bibfield  {journal} {\bibinfo  {journal} {Nucl. Instrum. Meth. A}\ }\textbf {\bibinfo {volume} {506}},\ \bibinfo {pages} {250} (\bibinfo {year} {2003})}\BibitemShut {NoStop}%
\bibitem [{\citenamefont {{GEANT4 Collaboration}}()}]{GEANT4:PhysicsManual}%
  \BibitemOpen
  \bibfield  {author} {\bibinfo {author} {\bibnamefont {{GEANT4 Collaboration}}},\ }\href {https://indico.cern.ch/event/679723/contributions/2792554/attachments/1559217/2454299/PhysicsReferenceManual.pdf} {\bibinfo {title} {{Physics Reference Manual}}},\ \bibinfo {note} {release 11.4, Rev11.4: December 5th, 2025}\BibitemShut {NoStop}%
\bibitem [{\citenamefont {{SBND Collaboration}}(2026)}]{sbndcode}%
  \BibitemOpen
  \bibfield  {author} {\bibinfo {author} {\bibnamefont {{SBND Collaboration}}},\ }\href@noop {} {\bibinfo {title} {{sbndcode: SBND LArSoft-based software framework}}},\ \bibinfo {howpublished} {\url{https://github.com/SBNSoftware/sbndcode}} (\bibinfo {year} {2026})\BibitemShut {NoStop}%
\bibitem [{\citenamefont {Hoecker}\ \emph {et~al.}(2009)\citenamefont {Hoecker} \emph {et~al.}}]{hoecker2009tmvatoolkitmultivariate}%
  \BibitemOpen
  \bibfield  {author} {\bibinfo {author} {\bibfnamefont {A.}~\bibnamefont {Hoecker}} \emph {et~al.},\ }\bibfield  {title} {\bibinfo {title} {{{TMVA} - Toolkit for Multivariate Data Analysis}},\ }\href {https://arxiv.org/abs/physics/0703039} {\  (\bibinfo {year} {2009})},\ \Eprint {https://arxiv.org/abs/physics/0703039} {arXiv:physics/0703039 [physics.data-an]} \BibitemShut {NoStop}%
\bibitem [{\citenamefont {Aguilar-Arevalo}\ \emph {et~al.}(2009{\natexlab{b}})\citenamefont {Aguilar-Arevalo} \emph {et~al.}}]{PhysRevLett.102.101802}%
  \BibitemOpen
  \bibfield  {author} {\bibinfo {author} {\bibfnamefont {A.~A.}\ \bibnamefont {Aguilar-Arevalo}} \emph {et~al.} (\bibinfo {collaboration} {MiniBooNE Collaboration}),\ }\bibfield  {title} {\bibinfo {title} {{Unexplained Excess of Electronlike Events from a 1-GeV Neutrino Beam}},\ }\href {https://doi.org/10.1103/PhysRevLett.102.101802} {\bibfield  {journal} {\bibinfo  {journal} {Phys. Rev. Lett.}\ }\textbf {\bibinfo {volume} {102}},\ \bibinfo {pages} {101802} (\bibinfo {year} {2009}{\natexlab{b}})}\BibitemShut {NoStop}%
\bibitem [{\citenamefont {Andreopoulos}\ \emph {et~al.}(2015)\citenamefont {Andreopoulos} \emph {et~al.}}]{Andreopoulos:2015wxa}%
  \BibitemOpen
  \bibfield  {author} {\bibinfo {author} {\bibfnamefont {C.}~\bibnamefont {Andreopoulos}} \emph {et~al.},\ }\href@noop {} {\bibinfo {title} {{The GENIE Neutrino Monte Carlo Generator: Physics and User Manual}}} (\bibinfo {year} {2015}),\ \Eprint {https://arxiv.org/abs/1510.05494} {arXiv:1510.05494 [hep-ph]} \BibitemShut {NoStop}%
\bibitem [{\citenamefont {Junk}(1999)}]{JUNK1999435}%
  \BibitemOpen
  \bibfield  {author} {\bibinfo {author} {\bibfnamefont {T.}~\bibnamefont {Junk}},\ }\bibfield  {title} {\bibinfo {title} {Confidence level computation for combining searches with small statistics},\ }\href {https://doi.org/https://doi.org/10.1016/S0168-9002(99)00498-2} {\bibfield  {journal} {\bibinfo  {journal} {Nucl. Instrum. Meth. A}\ }\textbf {\bibinfo {volume} {434}},\ \bibinfo {pages} {435} (\bibinfo {year} {1999})}\BibitemShut {NoStop}%
\bibitem [{\citenamefont {Read}(2002)}]{ALRead_2002}%
  \BibitemOpen
  \bibfield  {author} {\bibinfo {author} {\bibfnamefont {A.~L.}\ \bibnamefont {Read}},\ }\bibfield  {title} {\bibinfo {title} {{Presentation of search results: the CLs technique}},\ }\href {https://doi.org/10.1088/0954-3899/28/10/313} {\bibfield  {journal} {\bibinfo  {journal} {J. Phys. G}\ }\textbf {\bibinfo {volume} {28}},\ \bibinfo {pages} {2693} (\bibinfo {year} {2002})}\BibitemShut {NoStop}%
\bibitem [{\citenamefont {Fisher}(2006)}]{Fisher:2006zz}%
  \BibitemOpen
  \bibfield  {author} {\bibinfo {author} {\bibfnamefont {W.}~\bibnamefont {Fisher}} (\bibinfo {collaboration} {D0}),\ }\href@noop {} {\emph {\bibinfo {title} {{Systematics and limit calculations}}}},\ \bibinfo {type} {Tech. Rep.}\ \bibinfo {number} {\href{https://doi.org/10.2172/923070}{FERMILAB-TM-2386-E, D0-NOTE-5309}}\ (\bibinfo  {institution} {Fermilab},\ \bibinfo {year} {2006})\BibitemShut {NoStop}%
\bibitem [{\citenamefont {Heinrich}\ \emph {et~al.}(2021)\citenamefont {Heinrich}, \citenamefont {Feickert}, \citenamefont {Stark},\ and\ \citenamefont {Cranmer}}]{pyhf_joss_2021}%
  \BibitemOpen
  \bibfield  {author} {\bibinfo {author} {\bibfnamefont {L.}~\bibnamefont {Heinrich}}, \bibinfo {author} {\bibfnamefont {M.}~\bibnamefont {Feickert}}, \bibinfo {author} {\bibfnamefont {G.}~\bibnamefont {Stark}},\ and\ \bibinfo {author} {\bibfnamefont {K.}~\bibnamefont {Cranmer}},\ }\bibfield  {title} {\bibinfo {title} {pyhf: pure-python implementation of histfactory statistical models},\ }\href {https://doi.org/10.21105/joss.02823} {\bibfield  {journal} {\bibinfo  {journal} {J. Open Source Softw.}\ }\textbf {\bibinfo {volume} {6}},\ \bibinfo {pages} {2823} (\bibinfo {year} {2021})}\BibitemShut {NoStop}%
\bibitem [{\citenamefont {Cowan}\ \emph {et~al.}(2011)\citenamefont {Cowan}, \citenamefont {Cranmer}, \citenamefont {Gross},\ and\ \citenamefont {Vitells}}]{cowan_asymptotic_2011}%
  \BibitemOpen
  \bibfield  {author} {\bibinfo {author} {\bibfnamefont {G.}~\bibnamefont {Cowan}}, \bibinfo {author} {\bibfnamefont {K.}~\bibnamefont {Cranmer}}, \bibinfo {author} {\bibfnamefont {E.}~\bibnamefont {Gross}},\ and\ \bibinfo {author} {\bibfnamefont {O.}~\bibnamefont {Vitells}},\ }\bibfield  {title} {\bibinfo {title} {Asymptotic formulae for likelihood-based tests of new physics},\ }\href {https://doi.org/10.1140/epjc/s10052-011-1554-0} {\bibfield  {journal} {\bibinfo  {journal} {Eur. Phys. J. C}\ }\textbf {\bibinfo {volume} {71}},\ \bibinfo {pages} {1554} (\bibinfo {year} {2011})}\BibitemShut {NoStop}%
\bibitem [{\citenamefont {Pearson}(1901)}]{pearson1901pca}%
  \BibitemOpen
  \bibfield  {author} {\bibinfo {author} {\bibfnamefont {K.}~\bibnamefont {Pearson}},\ }\bibfield  {title} {\bibinfo {title} {{On Lines and Planes of Closest Fit to Systems of Points in Space}},\ }\href {https://doi.org/10.1080/14786440109462720} {\bibfield  {journal} {\bibinfo  {journal} {Lond. Edinb. Dubl. Phil. Mag. J. Sci.}\ }\textbf {\bibinfo {volume} {2}},\ \bibinfo {pages} {559} (\bibinfo {year} {1901})}\BibitemShut {NoStop}%
\bibitem [{\citenamefont {Jolliffe}(2002)}]{jolliffe2002pca}%
  \BibitemOpen
  \bibfield  {author} {\bibinfo {author} {\bibfnamefont {I.~T.}\ \bibnamefont {Jolliffe}},\ }\href@noop {} {\emph {\bibinfo {title} {{Principal Component Analysis}}}},\ \bibinfo {edition} {2nd}\ ed.,\ Springer Series in Statistics\ (\bibinfo  {publisher} {\href{https://doi.org/10.1007/b98835}{Springer}},\ \bibinfo {year} {2002})\BibitemShut {NoStop}%
\bibitem [{\citenamefont {Abdullahi}\ \emph {et~al.}(2025)\citenamefont {Abdullahi}, \citenamefont {Hoefken~Zink}, \citenamefont {Hostert}, \citenamefont {Massaro},\ and\ \citenamefont {Pascoli}}]{Abdullahi:2023ejc}%
  \BibitemOpen
  \bibfield  {author} {\bibinfo {author} {\bibfnamefont {A.~M.}\ \bibnamefont {Abdullahi}}, \bibinfo {author} {\bibfnamefont {J.}~\bibnamefont {Hoefken~Zink}}, \bibinfo {author} {\bibfnamefont {M.}~\bibnamefont {Hostert}}, \bibinfo {author} {\bibfnamefont {D.}~\bibnamefont {Massaro}},\ and\ \bibinfo {author} {\bibfnamefont {S.}~\bibnamefont {Pascoli}},\ }\bibfield  {title} {\bibinfo {title} {{Panorama of new-physics explanations to the MiniBooNE excess}},\ }\href {https://doi.org/10.1103/PhysRevD.111.035028} {\bibfield  {journal} {\bibinfo  {journal} {Phys. Rev. D}\ }\textbf {\bibinfo {volume} {111}},\ \bibinfo {pages} {035028} (\bibinfo {year} {2025})},\ \Eprint {https://arxiv.org/abs/2308.02543} {arXiv:2308.02543 [hep-ph]} \BibitemShut {NoStop}%
\bibitem [{\citenamefont {Abdullahi}\ \emph {et~al.}(2026)\citenamefont {Abdullahi} \emph {et~al.}}]{MicroBooNE:2025khi}%
  \BibitemOpen
  \bibfield  {author} {\bibinfo {author} {\bibfnamefont {A.~M.}\ \bibnamefont {Abdullahi}} \emph {et~al.} (\bibinfo {collaboration} {{MicroBooNE}}),\ }\bibfield  {title} {\bibinfo {title} {{First Search for Dark Sector ${e}^{+}{e}^{\ensuremath{-}}$ Explanations of the MiniBooNE Anomaly at MicroBooNE}},\ }\href {https://doi.org/10.1103/3q7x-ks7h} {\bibfield  {journal} {\bibinfo  {journal} {Phys. Rev. Lett.}\ }\textbf {\bibinfo {volume} {136}},\ \bibinfo {pages} {121804} (\bibinfo {year} {2026})}\BibitemShut {NoStop}%
\end{thebibliography}%
%################################################

\end{document}